\documentclass[apj]{emulateapj}

\usepackage{graphicx}
\usepackage{rotating}
\usepackage{afterpage}
\usepackage{textcomp}
\usepackage{txfonts}
\usepackage[]{natbib}
\bibpunct{(}{)}{;}{a}{}{,}

\newcommand{\um}{$\mu$m}

\newcommand{\kms}{km\thinspace s$^{-1}$}

\def\degr{\hbox{$^\circ$}}

\def\arcsec{\hbox{$^{\prime\prime}$}}
\def\utw{\smash{\rlap{\lower5pt\hbox{$\sim$}}}}
\def\udtw{\smash{\rlap{\lower6pt\hbox{$\approx$}}}}

\def\Lsun{\hbox{\it L$_\odot$}}

\def\Teff{\hbox{\it T$_{\rm eff}$}}

\def\Msun{\hbox{\it M$_\odot$}}

\def\Mbol{\hbox{\it M$_{\rm bol}$}}

\def\Teff{\hbox{\it T$_{\rm eff}$}}

\def\Mk{\hbox{\it M$_{\rm K}$}}
\newcommand{\Ks}{{\it K$_{\rm s}$}}

\newcommand{\Aks}{{\it A$_{ K_{\rm s}}$}}

\def\BCK{\hbox{\it BC$_{ K}$}}
\def\BCKs{\hbox{\it BC$_{ K_{\rm S}}$}}

\def\simgr{\mathrel{\hbox{\rlap{\hbox{\lower4pt\hbox{$\sim$}}}\hbox{$>$}}}}

\def\Vel{\hbox{\it Vel}}
\shorttitle{Late-type stars of class I  in Gaia DR2.}
\shortauthors{Messineo et al.}

\begin{document}


\title{A catalog of known Galactic  K-M stars of class I, candidate RSGs,  in Gaia DR2.
\footnote{\bf The full tables with 889 entries will only be available in the online Journal edition.}
}


\author{M.~Messineo\altaffilmark{1},
  A.G.A.~Brown\altaffilmark{2}
}

\altaffiltext{1}{Key Laboratory for Researches in Galaxies and Cosmology, University of 
Science and Technology of China, Chinese Academy of Sciences, Hefei, Anhui, 230026, China
\email{messineo@ustc.edu.cn}
}

\altaffiltext{2}{Leiden Observatory, Leiden University, Niels Bohrweg 2, 2333 CA Leiden, The Netherlands}

\begin{abstract}
We investigate individual distances and luminosities of a sample of  889 nearby candidate red
supergiants with reliable parallaxes ($\varpi / \sigma_\varpi > 4$ and $\text{RUWE} <2.7$) from Gaia
DR2. The sample was extracted from the historical compilation of spectroscopically derived spectral
types by \citet{skiff16}, and consists of  K-M stars  
that are listed   with class I  at least once.
The sample includes well-known red supergiants from \citet{humphreys78}, \citet{elias85},
\citet{jura90}, and \citet{levesque05}. Infrared and optical measurements from the 2MASS, CIO, MSX,
WISE, MIPSGAL, GLIMPSE, and NOMAD catalogs allow us to estimate the stellar bolometric magnitudes.
We analyze the stars in the luminosity versus effective temperature plane and confirm that 
43 sources are highly-probably red supergiants with \Mbol $< -7.1$ mag. 
 43\% of the sample  is made of  stars with masses  $>  7$ \Msun. Another $\approx$30\% of the 
sample consists of giant stars.
\end{abstract}


\section{Introduction}

The Milky Way is the closest laboratory for resolved stellar populations and a prototype of spiral
galaxies. Nonetheless our position within the disk and dust obscuration render its study difficult.
Red supergiants (RSGs) are the brightest stars seen at infrared wavelengths, being young and cold
objects with typical luminosity above $10^4$~\Lsun. RSGs are tracers of stellar populations from 4
to 30 Myr, with masses from about 9 to 40 \Msun\
\citep[e.g.][]{ekstroem12,chieffi13}; from their numbers and luminosities one can 
evaluate Galactic star formation in this range of time. The distribution of known spectral 
types of Galactic RSGs peaks at spectral types M0--M2 \citep{elias85,davies07}.

Having said that, the current census of RSGs, including the M-types is highly incomplete, 
with little being known about their spatial distribution 
\citep[see for example,][]{davies09,messineo16}. 
At optical wavelengths, catalogs of RSGs have been compiled by locating bright late-type stars 
in directions of OB associations. 
 \citet[][]{humphreys78} lists 92 RSGs, \citet{elias85} list 90 RSGs, 
\citet{levesque05} analysed the spectra of 62 RSGs,
\citet{jura90} list $\approx 135$ RSGs.
\citet{gehrz89} predicts at least 5000 RSGs. 
Overall, less than a thousand Galactic   late-type
stars of class I  are known, with only about 400  RSGs.
Their detection is extremely difficult since their
colors are similar to those of giant late-type stars and knowledge on their distances is poor,
and because their colours and magnitudes overlap with those of the more numerous 
Asymptotic giant branch (AGB) stars (from low masses to Super-AGBs of 9-10 \Msun).
Furthermore, even though associations and clusters make it easier to detect massive stars, it
appears that only $\approx 2$\% of inner Galaxy supergiants are associated with stellar clusters
\citep{messineo17}.
Pulsation properties and chemical abundances are required for identifying
the stage of evolution and the nuclear burning that has occurred.

Gaia data allows us to classify individual stars by providing their distances. 
We prepared a catalog of bright late-type stars reported at least 
once with class I, i.e., as stars of K- or M-type
and luminosity class I in the spectroscopic catalog of \citet[][]{skiff16}, and with data from Gaia DR2.
Historical spectroscopic records provided spectral types that in combination with Gaia parallaxes
and photometric data enabled us to measure the stellar luminosities. With that in hand, we were able
to extract a catalog of genuine stars  of luminosity class I
and to derive average magnitudes per spectral type. In
Sect.~\ref{data}, we describe the sample, their parallaxes, and available infrared measurements. In
Sect.~\ref{luminosity}, we estimate the stellar luminosities and provide average values per spectral
type. In Sect.~\ref{summary}, we summarize the results of our exercise.

\section{Observational data}
\label{data}

\subsection{The sample and available spectral types}
\label{catalogs}

We compiled a list of about 1400  K-M stars of class I with latitudes $|b|<10^\circ$ from the
historical records of stellar spectral types by \citet{skiff16}. 
All late-type stars with at least one classification as luminosity class I
were retained. 
In addition, we cross-matched Skiff's list with  existing Galactic compilations
of RSGs, for example by \citet{humphreys78}, \citet{elias85}, \citet{kleinmann86}, \citet{jura90},
\citet{caron03}, \citet{levesque05}, \citet{figer06}, \citet{davies08}, and
\citet{verhoelst09}.
 We also made use of the recent Galactic spectroscopic catalogues of bright late-type stars
by  \citet{blum03}, \citet{comeron04}, \citet{clark09}, \citet{liermann09}, \citet{rayner09}, 
\citet{negueruela10}, \citet{negueruela11}, \citet{verheyen12}, \citet{negueruela16},
\citet{messineo17}, and \citet{dorda18}.  Sources with available spectral types and good
parallaxes (see Sect. \ref{availableparallaxes}) are listed in Table \ref{table.gaia}. 
For  sources  listed in these recent catalogs, spectral classifications provided 
in the corresponding papers have been retained
(see footnotes to Table \ref{table.gaia}). The catalog by \citet{skiff16} collected spectroscopic
classifications of Galactic stars available from the literature, with some entries dating back to
1930--1950. 
For each star from one to a dozen entries were  available. For stars for 
which only one reference is  given (that to  Skiff's
database) we listed a spectral type range as well as the adopted spectral type, which is the mean
(or most recent) of the measured spectral types.

\subsection{Available parallaxes}
\label{availableparallaxes}

Gaia data were taken from the recently released Gaia DR2 catalog \citep{gaia2, gaiamission}, which
contains $1.7$ billion sources. Typically, for parallaxes of stars brighter
than $G=14$ mag, quoted uncertainties are about $0.04$ mas, $\approx 0.1$ mas for $G=17$ mag and
$\approx 0.7$ mas for $G=20$ mag (see \citet{gaiaplx2}).
Luminous late-type stars  are characterised by brightness fluctuations
due to convective motions and pulsation. The photocenters do not correspond
to the stellar barycenters, but fluctuate around it \citep[e.g.,][]{chiavassa11, pasquato11}.
This motion in general does not lead to systematic parallax errors, 
however, it degrades the goodness of fit of the astrometric solution \citep{chiavassa11}. 

Initial celestial positions were taken from the catalog of \citet{skiff16} and SIMBAD \citep{simbad}
and improved with the positions of available 2MASS matches. Gaia matches were searched using a
radius of 1\farcs 5. This resulted in 1342 Gaia sources, providing matches for 96\% of the initial
sample of late-type stars.  

For 7.5\% of the sample parallaxes were available from both the Gaia DR2 and Hipparcos catalogs
\citep{hip97}; the mean difference of parallaxes is $0.08$ mas, with a dispersion around the mean of
$1.21$ mas for stars with Gaia parallaxes larger than 2 mas.

\subsubsection{Astrometric quality filtering and best sample}

The goal of this work is to build a catalog of  secure known  K-M stars of class I, 
candidate RSGs, in Gaia DR2, and therefore to derive their 
average absolute magnitude for each spectral type. 
This means that here we  calculate the luminosity
of the candidate RSGs by direct integration of their stellar energy distribution (SED),
independently of colours or other information that might be obtained from the spectral energy
distribution. Hence, we rely on the Gaia DR2 parallax only to estimate the distances of the sources
in our sample. In order to make sure the corresponding luminosity  estimates are robust we will
apply a rather conservative filtering on the quality of the parallax data, as described in the
following.

Throughout the text we indicate with $\sigma_\varpi$ the external error of 
the parallax\footnote{\url{https://www.cosmos.esa.int/web/gaia/dr2-known-issues}},
which is defined as ${\sigma_\varpi}({\rm ext}) = \sqrt{k^2 \times {\sigma_\varpi}({\rm int})^2 + \sigma_{s}^2}$, where
${\sigma_\varpi}({\rm int})$ is the internal error provided by DR2, and $k = 1.08$ and  $\sigma_s = 0.021$ mas 
for $G < 13$ mag (bright), and $k = 1.08$ and $\sigma_s = 0.043$ mas for $G \ga 13$ (faint).

\begin{figure}
  \includegraphics[height=0.99\hsize]{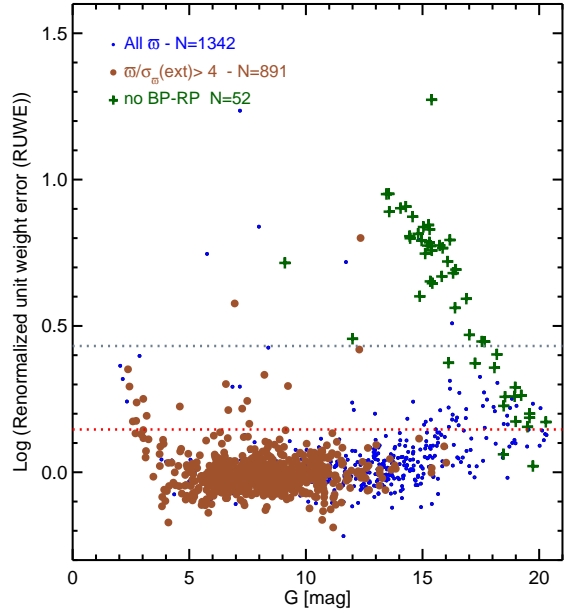}
  \caption{\label{fig:ruwe}
    The value of the \text{RUWE} vs.\ the apparent brightness in $G$ for all the source in our sample. The
    two lines indicate the limits $\text{RUWE}=1.4$ (in red) and $\text{RUWE}=2.7$ (in gray). The large brown dots
    indicate stars for which $\varpi/\sigma_\varpi$(ext)$>4$, while the dark-green crosses indicate stars for
    which no colour information is available from Gaia DR2.}
\end{figure}

In order to select sources with good quality astrometry  we analysed the
$\varpi/\sigma_\varpi$ ratio and  the so-called
renormalized unit weight error (\text{RUWE}) which the Gaia team recommends to use instead of the
filtering on the unit weight error described in appendix C of \citet{lindegren18}. 
The \text{RUWE} can be calculated using lookup tables available from the 
ESA Gaia web pages\footnote{\url{https://www.cosmos.esa.int/web/gaia/dr2-known-issues}} 
and it is described in detail
in a publicly available technical note \citep{lindegren18}. In Fig.~\ref{fig:ruwe} we show the
\text{RUWE} as a function of $G$ for all the sources in our sample.

Stars for which  $\varpi/\sigma_\varpi>4$  are indicated separately as well as stars for which 
no colour information is available (for which the value of the \text{RUWE} is less certain, 
this concerns 52 out of the 1342 sources in the sample). 
From this figure it is clear that  most sources for which
 $\varpi/\sigma_\varpi>4$  have a \text{RUWE} value below $1.4$ \citep[the threshold 
value recommended in][]{lindegren18}. A few stars with high signal to noise parallax values 
are located at $1.4<\text{RUWE}<2.7$. 
This suggests that a more relaxed filtering at $\text{RUWE}<2.7$
 is adeguate for RSGs, so as to retain the brightest stars 
for which the RUWE values may be affected by photocenter motions.

We further restricted our sample to stars with  $\varpi/\sigma_\varpi>4$ in order to ensure robust
distance estimates. We motivate this in the next section. In the end we thus retained 889 sources
with $\varpi/\sigma_\varpi > 4$ and $\text{RUWE}<2.7$. 
The parallax range of the sources after filtering is $0.19$ to $7.53$ mas.

\subsection{Distance estimates}

\begin{figure}
\resizebox{0.99\hsize}{!}{\includegraphics[angle=0]{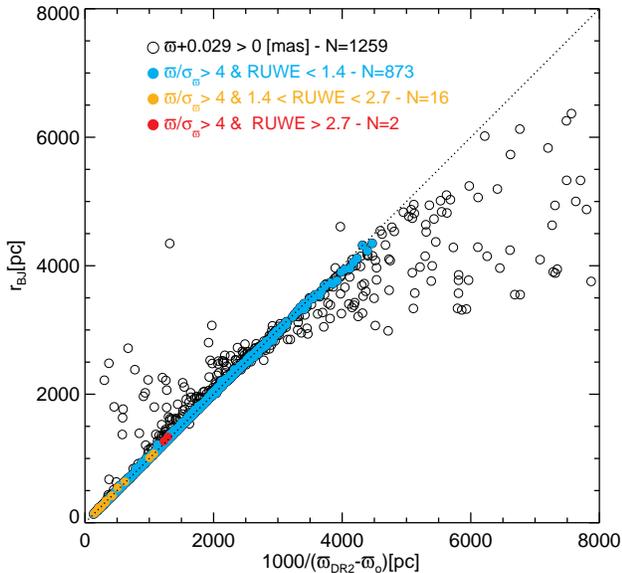}}
\caption{ \label{bias}  Gaia data. Parallactic distances inferred with the Milky Way model by 
\citet{bailer18} versus parallactic distances from direct inversion of the parallaxes.
Filled dots mark datapoints with $\varpi / {\sigma_\varpi}({\rm ext})  > 4$;
in cyan those  with $RUWE$ $< 1.4$, in orange $1.4 <$ $RUWE$ $\la 2.7$, and in red
$RUWE$ $> 2.7$. The dotted line shows the points of the equation
$r_{BJ} -( 1000/(\varpi- \varpi _o))=0$ pc.
}
\end{figure}

The proper use of parallaxes in the distance estimation problem has been extensively reviewed in the
context of Gaia DR2 by \cite{gaiaplx2}. Their recommendation is not to use the inverse of the
parallax as a distance indicator but to combine the parallax with other information and treat the
estimation of distance as an inference problem. In our case we wish to use only the parallax in
order to establish the luminosity  of our stars independent from other information and in that
case the Bayesian distance estimation method proposed by \cite{bailerjones15}, in particular using
the exponentially decreasing space density prior, is a good choice \citep{gaiaplx2}. We will use the
distances estimated by \citet{bailer18} for our selection of source with good quality and precise
parallaxes, and motivate this as follows. For parallaxes with $\varpi/{\sigma_\varpi}({\rm ext})>4$ the
\cite{bailer18} distances {\em by design} give essentially the same result as the $1/\varpi$
estimator, because for any reasonable length scale, $L$, of the exponentially decreasing space density
prior the likelihood dominates the posterior on the distances. At larger relative parallax error, the
prior plays a stronger role which would make our luminosity class estimates somewhat dependent on
the Galactic model employed as a prior by \cite{bailer18}. We verified that for our sources the
relative differences between the $1/(\varpi-\varpi_0)$ \footnote{$\varpi_0=-0.029$~mas is the parallax
zero point estimated by \citet{lindegren18}} and \cite{bailer18} distance estimates are less than 5
per cent  (see Fig.\ \ref{bias}), with no trends as a function of the value of $L$. 
A summary of relative differences between the $1/(\varpi-\varpi_0)$ and the Bailer's distances ($R_{BJ})$
are provided in Table \ref{chifracchia}.

\begin{sidewaystable*} [t]
\caption{\label{chifracchia} Average  difference of the distances  provided  by \citet{bailer18}, $<R_{BJ}>$ 
and  distances from direct inversion of the parallaxes for stars 
with $\varpi/{\sigma_\varpi}({\rm ext}) >$ 2,3,4,5, and 10.}
\begin{tabular} {r r | r r r r r r| r r r r r r r}
\hline
\hline
                        &                 &\multicolumn{6}{c}{ All } &\multicolumn{6}{c}{$dist > 3.5$ kpc }\\
$\varpi / \sigma_\varpi$& $N_{\rm stars}$ & $\Delta$(dist)& $\sigma$ & $<\Delta$ (M1)$>$ & $\sigma$& $<\Delta$ (M2)$>$ &$\sigma$&$\Delta$(dist) &$\sigma$ & $<\Delta$ (M1)$>$ &$\sigma$  &$<\Delta$ (M2)$>$ &$\sigma$  \\\\
                        &                 & [pc]           & [pc]     & [mag]           &     [mag] & [mag]             &[mag]   &         [pc]  & [pc]    & [mag]           &     [mag]  & [mag]            &[mag]        \\
\hline
2  &                    1075&                  $-$5.58 &  123.74  & 0.017 & 0.093    & 0.63 & 0.56 &$-$289.04  &    316.95 & $-0.13$ &   0.11 & $0.96$ & 0.25\\
3  &                    981&                  2.78    &  44.44   & 0.014 & 0.045    & 0.51 & 0.38 &$-$124.94  &    105.35 & $-0.06$ &   0.04 & $0.82$ & 0.18\\
{\bf 4 }                &{\bf 891   }               & {\bf 3.98}    & {\bf 22.42}   & {\bf 0.011} &{\bf 0.026 }   & {\bf 0.45} &{\bf 0.25} &$\bf -83.72$   & {\bf 47.32}& $\bf -0.05$ &  {\bf 0.03 }& $\bf 0.70$ & {\bf 0.08}\\
5  &  805 &5.22                               &  13.13   & 0.01 & 0.02      & 0.39 & 0.20 &$-$58.394  &     23.28 & $-0.03$ &   0.01 & $0.62$ & 0.01\\
10  & 379  & 2.92                              &   2.59   &  0.006 &  0.007  & 0.21 & 0.10 &   $..$      &   $..$  &                                     \\
\hline
\end{tabular}
\begin{list}{}
\item  {\bf Notes:} $\Delta$(dist)$=<R_{BJ}-1000/(\varpi-\varpi_0)>$.~\\
$<\Delta  (M1)>$ is the difference in the Distance Moduli inferred with the  two  distances $<R_{BJ}>$ and $<1000/(\varpi-\varpi_0)>$.~\\
$<\Delta  (M2)>$ is the difference in the Distance Moduli of the high and low distances inferred by \citet{bailer18}.\\
\end{list}
\end{sidewaystable*}

Using the distance estimates from \cite{bailer18} even for our sample with very precise parallaxes
has the added advantage that the uncertainties on the distance estimates (as well as on the distance
moduli used below) are well defined. On the contrary, the $1/\varpi$ distance estimator follows a
probability distribution which cannot be normalised and thus has no expectation value or variance.
Parallax uncertainties propagated into distance uncertainties
($\sigma_d\approx\sigma_\varpi/\varpi^2$) are thus formally meaningless for the $1/\varpi$ distance
estimator \citep[see][]{gaiaplx2}.

\begin{sidewaystable*} [b]
\caption{ \label{table.gaia} Parallaxes and spectral types of the  889 stars with $\varpi / \sigma_\varpi > 4$ and $\text{RUWE} < 2.7$.}
{\tiny
\renewcommand{\arraystretch}{0.6}
\begin{tabular}{@{\extracolsep{-.07in}}rlrr|rrrrrr|llr|rr|l}
\hline 
\hline 
   & & & &\multicolumn{6}{c}{\rm Gaia}   &   \multicolumn{3}{c}{\rm Sptype} &  \multicolumn{2}{c}{\rm Distance} &  \multicolumn{1}{c}{\rm Cluster} \\ 
\hline 
Id & Alias & {\rm Ra(J2000)} & {\rm Dec(J2000)} &{\rm ID}   & 
 {$\varpi$} & {\rm pmRa} & {\rm pmDec} & {\rm G} &   {\Vel$^*$} &
 {\rm Sp(Skiff)}  &  {\rm Sp(adopt)}  &{\rm Ref} & Inv & MW  &\\ 
\hline 
 & &    {\rm [hh mm ss]}    & {\rm [dd mm ss]} &    & { \rm [mas]}  & {\rm [mas yr$^{-1}$]}  & {\rm [mas yr$^{-1}$]}  & {\rm [mag]} &{\rm [\kms]} &
                  &                               &           &[pc]&[pc]&\\ 
\hline 

    1  &${\rm                         PER002    }$  &   0:00:18.123  &  60:21:01.538  &            423337510285997440  &  1.32$\pm$  0.07  & -6.831$\pm$  0.081  & -1.540$\pm$  0.089  &  6.784$\pm$  0.002  & $..$   &                                 $..$    &        M4.5        Ib       &                         4   &       743 &   744 $^{+   33 }_{-   31   }$ &                       $..$     \\
    2  &${\rm                         PER006    }$  &   0:02:59.105  &  61:22:05.344  &            429500547840721536  &  0.98$\pm$  0.04  & -1.181$\pm$  0.059  & -1.221$\pm$  0.056  &  8.490$\pm$  0.001  &  -45.580$\pm$    0.190  &                                 $..$    &          M3        Ib       &                         4   &       990 &   992 $^{+   36 }_{-   34   }$ &                       $..$     \\
    3  &${\rm                         PER008    }$  &   0:06:38.571  &  58:02:18.208  &            422677631507971840  &  0.82$\pm$  0.08  & -3.328$\pm$  0.099  & -3.282$\pm$  0.089  &  9.598$\pm$  0.002  & $..$   &                                 $..$    &          M5        Ib       &                         4   &      1176 &  1183 $^{+  117 }_{-   98   }$ &                       $..$     \\
    4  &${\rm                         PER010    }$  &   0:09:26.327  &  63:57:14.090  &            431678852171577216  &  0.40$\pm$  0.07  & -3.633$\pm$  0.098  & -0.372$\pm$  0.110  &  6.768$\pm$  0.012  &  -54.300$\pm$    0.530  &                                 $..$    &          M2       Iab       &                         4   &      2350 &  2355 $^{+  423 }_{-  314   }$ &                       $..$     \\
    5  &${\rm                         KN~Cas    }$  &   0:09:36.363  &  62:40:04.091  &            429999760479435520  &  0.29$\pm$  0.06  & -1.850$\pm$  0.077  & -1.817$\pm$  0.059  &  8.356$\pm$  0.002  & $..$   &                                 $..$    &          M1        Ib       &                     1,5,9   &      3131 &  3082 $^{+  558 }_{-  416   }$ &                  Cas~OB5   \\
    6  &${\rm                         PER012    }$  &   0:12:21.655  &  62:53:33.738  &            431331097263392384  &  0.95$\pm$  0.04  & -1.455$\pm$  0.043  & -2.351$\pm$  0.044  &  6.914$\pm$  0.000  &  -35.120$\pm$    0.150  &                                 $..$    &          K0       Iab       &                       1,4   &      1025 &  1026 $^{+   29 }_{-   27   }$ &                       $..$     \\
    7  &${\rm                         PER015    }$  &   0:15:01.100  &  66:06:50.122  &            528168213046737024  &  2.16$\pm$  0.04  &  5.334$\pm$  0.048  & -5.527$\pm$  0.046  &  7.231$\pm$  0.001  &  -32.050$\pm$    0.180  &                                 $..$    &          K3        Ib       &                         4   &       456 &   456 $^{+    7 }_{-    7   }$ &                       $..$     \\
    8  &${\rm                         PER019    }$  &   0:18:26.380  &  60:54:09.149  &            428817510598195584  &  0.42$\pm$  0.04  & -2.826$\pm$  0.049  & -1.200$\pm$  0.044  &  7.795$\pm$  0.001  &  -49.280$\pm$    0.170  &                                 $..$    &          M1       Iab       &                         4   &      2222 &  2220 $^{+  165 }_{-  144   }$ &                       $..$     \\
    9  &${\rm                         PER022    }$  &   0:20:43.560  &  61:52:46.537  &            430464235421496320  &  0.81$\pm$  0.09  & -1.599$\pm$  0.104  & -0.334$\pm$  0.094  &  5.760$\pm$  0.002  &  -29.740$\pm$    0.320  &                                 $..$    &          M1       Iab       &                       4,8   &      1188 &  1198 $^{+  130 }_{-  107   }$ &                       $..$     \\
   10  &${\rm                      BD~+59~38    }$  &   0:21:24.278  &  59:57:11.155  &            428379733171150336  &  0.53$\pm$  0.07  & -3.470$\pm$  0.084  & -0.924$\pm$  0.070  &  7.966$\pm$  0.005  &  -55.570$\pm$    0.850  &                                 $..$    &       M2/M2     Iab/I       &                 1,2,5,8,9   &      1778 &  1783 $^{+  230 }_{-  184   }$ &                  Cas~OB4   \\
\hline
\end{tabular}
}
\begin{list}{}
\item {\bf Notes:}  The identification number (Id) is followed by an Alias name, the Gaia coordinates,
the Gaia parameters (name=ID, parallax=$\varpi$ and  its external error ($\sigma_\varpi$), proper motions, $G$-band magnitude, \Vel),
the spectral types (Sp(Skiff)) collected by \citet{skiff16}, 
the adopted spectral type (Sp(adopt)), references for the spectral types  (Ref), distances, and nearby clusters.\\
Sp(adopt) is that of the first reference listed which is $\ne 1$. 
When only Skiff's reference is present (=1), an average spectral type from Skiff's records is adopted 
and the encountered spectral range is annotated (Sp(Skiff)).
When \citet{levesque05} reference is present (=2), two values are provided, the photographic MK type and class, and
the new type by \citet{levesque05} (revised by  fitting synthetic models).\\
"Inv" distances are obtained by inversion of the parallaxes, "MW" distances and relative errors
are those of \citet{bailer18}, and are based on a prior derived from a Milky Way model.\\
 ($^*$) Spectroscopic radial velocity in the solar barycentric reference frame.\\

\item {\bf References:} 
2=\citet{levesque05};  
3= \citet{verhoelst09};  
4=\citet{dorda18}; 
5= \citet{negueruela16}; 
6=\citet{kleinmann86}; 
7=\citet{elias85}; 
8=\citet{jura90};  
9=\citet{humphreys78}; 
10=\citet{messineo17}; 
11=\citet{messineo14b}; 
12=\citet{negueruela12};  
13=\citet{negueruela11}; 
14=\citet{rayner09};      
15=\citet{liermann09};    
16=\citet{mermilliod08};  
17=\citet{messineo08};    
18=\citet{mengel07};      
19=\citet{caron03};       
20=\citet{massey01};      
21=\citet{eggenberger02}.~\\ 
\end{list}
\end{sidewaystable*}

\subsubsection{RSGs related to clusters and radio parallaxes}

In this work, we treated the stars individually. However, in Table \ref{table.gaia} we have
annotated possible associations with known clusters, which is based on current literature. Only
13\% of the sample was found associated. Memberships are not the focus of this work as they require
an extensive and careful revision of each open cluster. For example, with Gaia DR2 data doubt is
cast even upon the association of $\eta$ Car with the young cluster Trumpler 16 \citep{davidson18}.

The 22 RSGs reported in Table \ref{table.gaia} as associated to the Per OB1 association yield an
average $\varpi+0.029=0.51$ mas with a dispersion around the mean of $0.13$ mas, or an average
$\varpi+0.029=0.54$ mas with a dispersion of $0.11$ mas when including only the best quality
sources. The annual parallax of maser spots measured toward S Persei is $0.413 \pm 0.017$ mas
\citep{asaki10}. Unfortunately, the Gaia parallax of S Persei  (G=7.80 mag)	has 
a large uncertainty, $\varpi=0.22 \pm  0.13$ mas, RUWE=1.27, $\varpi/{\sigma_\varpi}({\rm ext})$=1.67.

\citet{zhang12a} and \citet{choi08} reported 
on astrometric observations of $H_2O$ masers around the red supergiant VY Canis Majoris ($G=7.17$ mag).
The trigonometric parallax is $0.88\pm0.08$ mas, corresponding to a distance of $1.14^{0.11}_{-0.09}$ kpc.
Unfortunately, Gaia measurements are highly uncertain  ($\varpi=-5.92\pm    0.89$ mas, RUWE=17.19).

The red hypergiant VX Sgr ($G=7.17$ mag) has a  trigonometric parallax of 
$0.64 \pm 0.04$ mas, corresponding to a distance of $1.56_{-0.10}^{+0.11}$ kpc \citep[via water maser observations,][]{xu18}.
\citet{chen07} had estimated a distance  of $1.57\pm0.27$ kpc with SiO maser observations.
Gaia parallax is $\varpi=0.79 \pm  0.27$ mas, $1.36^{+1.02}_{-0.41}$ kpc 
(RUWE=1.96, $\varpi/{\sigma_\varpi}({\rm ext})$=3.17).  
VX Sgr remains outside of our selected 889 stars because of its low $\varpi/\sigma_\varpi$,
however, the radio parallax and Gaia parallax agree within 23\%.

The red supergiant PZ Cas ($G=6.64$ mag) has an annual parallax of $0.356 \pm 0.026$ mas, 
corresponding to a distance of 
$2.81^{+0.22}_{-0.19}$ kpc \citep[from water maser observations,][]{kusuno13}. 
Gaia measurements are consistent within errors 
($\varpi=0.42\pm 0.09$ mas, $2.22^{+0.53}_{-0.36}$ kpc, RUWE=1.06, $\varpi/{\sigma_\varpi}({\rm ext})$=4.67).
PZ Cas is listed in Table \ref{table.gaia}. The radio and Gaia parallaxes agree within 18\%.

\subsection{Photometric catalog}

Photometric $JH$\Ks\ measurements from the Two Micron All Sky Survey (2MASS) catalog
\citep{skrutskie06,cutri03} were available for 97\% of the sample in Table \ref{table.gaia}. Their
\Ks\ values range from $-4$ mag to about 12.5 mag. Of the \Ks\ magnitudes 43\% are brighter than \Ks
$= 4$ mag, and magnitudes are based on the fitting of the wing of the PSF on the 51ms exposures (red
flag  Rk = 3, see Table \ref{table.gaia}). For 6.5\% of these stars, we were also able to retrieve $J$,
$H$, and $K$ measurements in the Catalog of Infrared Observations, CIO 5th edition, by
\citet{gezari96}; the average difference at 2 \um\ is $0.13$ mag with $\sigma = 0.14$ mag. For the
remaining 2.7\% of the sample with missing near-infrared measurements, we used the photometry of
\citet{morel78}, \citet{liermann09}, \citet{messineo10}, and \citet{stolte15}. For the faintest star
OGLE~BW3~V~93508 ($K=13.9$ mag) the measurements are from \citet{lucas08}.

For 78\% of the stars mid-infrared measurements from the Midcourse Space Experiment
\citep[MSX,][]{egan03,price01} were available. For 27\% of the sample 24 \um\ measurements from
MIPSGAL by \citet{gutermuth15} were available. For 32\% of the sample there were GLIMPSE
measurements \citep{churchwell09,benjamin05}; for 96\% mid-infrared measurements from 3.6 \um\ to
22 \um\ were available from the Wide-field Infrared Survey Explorer (WISE) \citep{wright10}. We
used an initial search radius of 5\arcsec\ and selected the closest matches. The MSX matches were
at an average distance of 1\farcs 3 with $\sigma$=0\farcs 9 from the 2MASS positions; the WISE
matches at an average distance of 0\farcs 4 ($\sigma$=0\farcs 4). The Gaia positions were searched
to within 1\farcs 5 of the 2MASS positions, and have an average displacement of 0\farcs 17 and a
$\sigma$=0\farcs 13 from the 2MASS centroids; 2MASS stars are the closest matches to the Gaia
sources and also the brightest \Ks\ sources. Matches were confirmed with a visual inspection of
2MASS and WISE images, as well as of the stellar  energy distribution (SED). 
Notes on the matches are provided in Appendix A.

$BVR$ photometry was retrieved from The Naval Observatory Merged Astrometric Dataset (NOMAD) 
\citep{zacharias05}. The photometric data for the subsample of 889 stars with 
good parallaxes are listed in Table \ref{table.magnitudes}.

\begin{table*} 
\begin{center}
{\tiny
\caption{\label{table.magnitudes} Infrared measurements of the bright late-type stars in Table \ref{table.gaia}.} 
\begin{tabular}{@{\extracolsep{-.07 in}} l|rrrrrrrrr|rrr|rrrr|rrrr|rrrr|r|rrr|rr|rr}
\hline 
 &   \multicolumn{9}{c}{\rm 2MASS$^*$} &   \multicolumn{3}{c}{\rm CIO} &   \multicolumn{4}{c}{\rm GLIMPSE}   & \multicolumn{4}{c}{\rm MSX}&  \multicolumn{4}{c}{\rm WISE}  & \multicolumn{1}{c}{\rm MIPS} & \multicolumn{3}{c}{\rm NOMAD} & Nstar$^+$ & \\ 
\hline 
 {\rm ID} &   {\it J} & Rj&Qj & {\it H} & Rh& Qh& { $K_S$} &Rk &Qk &
 {\it J} & {\it H} & {\it K} &
  {\rm [3.6]} & {\rm [4.5]} & {\rm [5.8]} & {\rm [8.0]} &
 {\it A}  & {\it C}  &{\it D}  &{\it E}  &
 {\it W1} &{\it W2}  & {\it W3} &  {\it W4} & {\rm [24]} & {\it B}& {\it V}& {\it R}& \\ 
\hline 
  &  1.2 & & &  1.6 & & & 2.2  & & &
  1.25 & 1.65 & 2.20 &
 3.6 & 4.5 & 5.8 & 8.0 &
 8.3  & 12.1  &14.6  &21.3  &
 3.4 &4.6  & 11.6 &  22.1 & 23.7 \\ 
 
\hline 
  &   {\rm [mag]}   & & & {\rm [mag]} & & & {\rm [mag]} & & & 
  {\rm [mag]}   &{\rm [mag]}   &{\rm [mag]}   &
{\rm [mag]} & {\rm [mag]}  & {\rm [mag]}&{\rm [mag]}&{\rm [mag]}&{\rm [mag]} &{\rm [mag]}&{\rm [mag]}&{\rm [mag]}& {\rm [mag]}& {\rm [mag]}& {\rm [mag]}&{\rm [mag]}&{\rm [mag]}&{\rm [mag]}& {\rm [mag]}&\\ 
\hline 
     1   &  3.56 & 3 & D &  2.64 & 3 & C &  2.18 & 3 & D & $..$ & $..$ & $..$ & $..$ & $..$ & $..$ & $..$ &  1.87 &  1.86 &  1.62 & $..$ & $..$ & $..$ &  1.97 &  1.79 & $..$& 10.15 &  8.48 &  7.60 &   110 &   \\
     2   &  5.53 & 1 & A &  4.62 & 1 & A &  4.30 & 1 & A & $..$ & $..$ & $..$ & $..$ & $..$ & $..$ & $..$ &  4.17 & $..$ & $..$ & $..$ &  4.19 &  4.05 &  4.21 &  4.06 & $..$& 11.49 &  9.75 &  8.87 &   110 &   \\
     3   &  5.81 & 1 & A &  4.86 & 1 & E &  4.48 & 1 & A & $..$ & $..$ & $..$ & $..$ & $..$ & $..$ & $..$ &  4.30 & $..$ & $..$ & $..$ &  4.37 &  4.32 &  4.29 &  4.11 & $..$& 16.46 & $..$ & 10.50 &   110 &   \\
     4   &  3.21 & 3 & D &  2.15 & 3 & D &  1.73 & 3 & D & $..$ & $..$ &  1.81 & $..$ & $..$ & $..$ & $..$ &  0.17 & -0.42 & -0.39 & -1.18 & $..$ & $..$ & -0.23 & -1.22 & $..$& 10.22 &  8.37 &  7.49 &   110 &   \\
     5   &  5.25 & 1 & A &  4.53 & 3 & D &  4.29 & 3 & D & $..$ & $..$ & $..$ & $..$ & $..$ & $..$ & $..$ &  3.74 &  3.60 & $..$ & $..$ &  3.80 &  3.68 &  3.73 &  3.53 & $..$& 11.30 &  9.57 &  8.69 &   110 &   \\
     6   &  5.03 & 3 & D &  4.08 & 3 & D &  3.64 & 1 & E & $..$ & $..$ & $..$ & $..$ & $..$ & $..$ & $..$ &  3.48 &  3.44 & $..$ & $..$ &  3.53 &  3.40 &  3.55 &  3.47 & $..$&  9.25 &  7.55 &  6.67 &   110 &   \\
     7   &  4.53 & 3 & D &  3.66 & 3 & C &  3.36 & 3 & D & $..$ & $..$ & $..$ & $..$ & $..$ & $..$ & $..$ &  3.27 &  3.39 & $..$ & $..$ & $..$ &  3.25 &  3.34 &  3.23 & $..$& 10.11 &  8.20 &  7.32 &   110 &   \\
     8   &  4.79 & 3 & D &  3.74 & 3 & D &  3.25 & 3 & D & $..$ & $..$ & $..$ & $..$ & $..$ & $..$ & $..$ &  2.98 &  2.77 &  2.69 & $..$ & $..$ &  3.06 &  2.99 &  2.60 & $..$& 11.41 &  9.10 &  8.53 &   110 &   \\
     9   &  3.12 & 3 & D &  2.26 & 3 & C &  1.88 & 3 & D & $..$ & $..$ &  1.75 & $..$ & $..$ & $..$ & $..$ &  1.53 &  1.44 &  1.37 &  1.45 & $..$ & $..$ &  1.65 &  1.48 & $..$&  8.82 &  6.85 &  5.97 &   110 &   \\
    10   &  4.58 & 3 & D &  3.43 & 3 & D &  2.71 & 3 & D & $..$ & $..$ & $..$ & $..$ & $..$ & $..$ & $..$ &  0.97 &  0.28 &  0.45 & -0.12 & $..$ & $..$ &  0.50 & -0.12 & $..$& 11.82 &  9.66 &  8.94 &   110 &   \\

\hline
\end{tabular}
}
\begin{list}{}
\item {\bf Notes:}
The identification number (Id) is followed by the 2MASS $JHK$ measurements with corresponding
red flags (Rj,Rh,Rk) and quality flags (Qj,Qh,Qk), CIO $JHK$ magnitudes, MSX $A,C,D,E$ magnitudes,
WISE $W1,W2,W3,W4$ magnitudes, MIPS 24 \um\ magnitude, the NOMAD $BVR$ magnitudes, and the Nstar value.~\\
($^+$) Nstar=XYZ, where X=number of MSX detected within the search radius; 
Y= number of WISE stars within the search radius; 
Z= number of GLIMPSE stars with 8 \um\ magnitudes $< 10$  within the search radius.
A value equals to 9 indicates that the counter is not available.~\\
($^*$)If the 2MASS quality flags are equal to 'M' the measurements have other origins
as specified in Appendix A.~\\
A few WISE and MSX measurements were discarded (Appendix A).
\end{list}

\end{center}
\end{table*}

\section{Luminosities}
\label{luminosity}

\subsection{Bolometric magnitudes}
\label{calcmbol}

We estimated the stellar luminosities using the photometric measurements, an extinction power law
with an index of 1.9 \citep[][]{messineo05}, and the distance moduli derived from the Gaia
parallaxes. For spectral types from K0 to M5, intrinsic $J-$\Ks\ and $H-$\Ks\ colours were taken
from \citet{koorneef83}. For M6--M9 types intrinsic colours were derived from the colours of giants
\cite[e.g.][]{koorneef83, montegriffo98, cordier07} and the average offset between the colours of
giants and supergiants of types M3--M5 were applied. Bolometric corrections to the absolute
$K$-magnitudes were provided by \citet{levesque05}. In addition to this calculation, we performed a
direct flux integration using the $JH$\Ks\ measurements, and the mid-infrared measurements from MSX,
WISE, GLIMPSE, and MIPSGAL. Measurements were dereddened with extinction ratios as described in
\citet{messineo05}. The integral under the stellar energy distribution (SED) was estimated with the
trapezium method; flux extrapolations at the red-extremes were performed with a linear interpolation
passing through the last reddest data-point and going to zero flux at 500 \um, while at the
blue-extreme (bluer than $J$-band) we use a black body extrapolation \citep[see ][]{messineo17}.
Red extrapolation contains about 5\textperthousand\ of the flux. 
The average difference between
the  \Mbol\ calculated with the \BCKs\ and those calculated by integrating under the SED 
is 0.05 mag with a $\sigma$=0.18 mag. 
Inferred \Mbol\ values are listed in Table \ref{table.properties}.

We estimated de-reddened $BV$ photometry, $V_{\rm o}$ and $B_{\rm o}$, by using the estimated \Aks\
and assuming $R=3.1$ and the extinction ratios in \citet{messineo05}.

\subsection{Luminosity classes and nuclear burnings}
\label{lumclass}

The MK system  was established in 1943 by Morgan and Keenan,
and it is an empirical system  for the stellar spectral classification. It is  based on a
known atlas of standard stars with   spectral types  and luminosity classes \citep{morgan43}.
Stellar spectra are classified by direct comparison with spectra of standard stars observed
at the same resolution and with the same instrument.  
Through quantitative spectral analysis one can estimate gravity, $g$, or  \Teff,
however, such
quantities are external to the definition of MK system itself.
While spectroscopic indicators of luminosity for dwarfs and evolved late-type stars are at our
disposal from atomic lines and molecular bands, the separation of giants and supergiants 
remains difficult.  Furthermore, spectroscopic optical and infrared classifications 
may provide somewhat different results \citep{gray09};  supplementary information on 
distances, luminosities,  and chemical composition is necessary.

Higher extinction renders the $M_{V}$ versus $B_{\rm o}-V_{\rm o}$ 
unsuitable for studies of the inner Galaxy, 
and it is useful to translate the optical quantities into infrared quantities 
and theoretical quantities.   Furthermore, it is useful to look at these diagrams
by keeping in mind which types of nuclear burnings may occur.

AGBs and RSGs are cold objects with similar ranges of effective temperatures, 
therefore spectral types. 
They overlap in luminosity. AGB stars can even be brighter  than RSGs,
and it is not known apriori from the luminosity classes 
the type of internal nuclear burnings and neither their distances.

AGB stars  are stars of low or intermediate masses ($\la 9$ Msun)
burning helium and hydrogen in shells, with a degenerate core of CO. 
AGB stars from 6.5 to 9.5 \Msun\  experience off-center nuclear burnings and
from 9 to 10 \Msun\ can even reach  iron core state and evolve into neutron star.

As \citet{iben74} writes,  massive stars are
{\it stars which do not develop a strongly electron-degenerate core until all 
exoergic reactions have run to completion at the center}.
RSGs are massive stars from $\approx 9$ to $\approx 40$ \Msun\ \citep[][]{ekstroem12}. 
Most of them are burning He when they reach the RSG phase.
For a RSG of 9 \Msun\  models predict \Mbol\ from $-4.5$ to $-6.8$ mag and 
spectral types from K0 to M4.5, while for a RSG of 25 \Msun,  \Mbol\  
$\approx -8.8$ mag and  spectral type  K5 (see Table \ref{bench}).
Observations closely follow the new evolutionary tracks by \citet[][]{ekstroem12}.
The \Mbol\ values of the $\approx 90$ Galactic RSGs recently analysed by \citet{levesque05} range
from \Mbol=$-3.63$ mag to \Mbol= $-10.36$ mag.

A few observational luminosity benchmarks of late-type stars 
 of low and intermediate masses are here useful. The tip
of red giant branch stars in Galactic globular clusters  occurs at 
\Mbol\ = $-3.6$ to $-3.8$ mag in metal-rich globular clusters, such us 47 Tuc
\citep[e.g.][]{ferraro00}; members brighter than that are thermally pulsing TP-AGBs. 
The maximum luminosity that more massive AGB stars can reach is about $-7.1$ mag \citep{wood93}.  
Very massive AGB stars may experience hot-bottom burning
which further increases their luminosity, 
but this phenomenon primarily affects metal poor populations and is thus expected to 
only moderately affect the Milky Way disk population.
The latest models of \citet{doherty15} predict that a super-AGB of 9 \Msun\ 
would reach \Mbol $= -7.6$ mag.
Therefore, AGBs do have a large overlap in luminosity with RSGs, and may
enter the  luminosity classes Ia, Ib, and Ib-II;
for example, as pointed out by the kind referee,
$\alpha$ Her is an AGB of 2--3 \Msun\ with class Ib-II \citep[][]{moravveji13},
and NGC6067 hosts several AGBs of 6 \Msun\ with types K0-K4 and classes 
Iab-Ib, Iab-Ib and Ib \citep[][]{alonsosantiago17}.

However, observationally, we can see that field AGB stars in the Baade's Windows  with \Mbol\ from $\approx-5.0$ to $-7.1$ mag are  
large amplitude pulsators (Miras) \citep[e.g.][]{alard01}, and generally have late-$M$ spectral types,
M4-M9 \citep[i.e. \Teff\ cooler than 3500 K,][]{alard01,blanco84};
similarly, the 4 Mira stars (V1-V4) at the tip of the red branch of the globular cluster 47 Tuc have
spectral types M4-M5 \citep{glass73,skiff16}. By contrast,  semiregular AGB pulsators are typically
fainter than Mira AGBs: $-2.5 \ga$ \Mbol $\ga -5.0$ mag, while Miras
have $-3.6 \ga$ \Mbol $\ga -7$ mag \citep[e.g.,][]{alard01}. 

In conclusion, only stars brighter than \Mbol$\approx -7.5$ mag  (masses $> 15$ \Msun) 
are certain RSGs; late-type stars earlier than $M4$ and with \Mbol $\la -5.0$ mag are expected to have
masses $\ga 5-7$ \Msun.
For field late-type stars fainter or redder than that,  
AGB stars are  the dominant  population when \Mbol$<-3.6$ mag (see Table \ref{bench}).

\begin{table*}
\caption{\label{bench} Summary of  \Mbol\ and temperatures of Galactic massive cool stars (RSGs)
and other cool stars of low and intermediate masses. } 
\begin{tabular}{rrllrllll}
\hline
\hline
Mass   & Age\_to\_red & T\_red &Phase & \Mbol          & \Teff        & Sp. Type &Comments\\
\Msun  & [Myr]      & [Myr]  &    & [mag]          &  [K]         &          &  \\
\hline
0.6-0.8&     & &tip-rgb  & [$-3.6$,$-3.8$]  &              &          & Observed range in globular clusters \citep{ferraro00}\\
1.35-1.7 &     & &tip-rgb  & [3.4]         &              &          & Rot. tracks  by \citet[][]{ekstroem12}\\
$<2.0-2.8$ &   & &tip-rgb  &  [$ -3.5$,$ -3.7$]  &              &          & He-flash theory  for Z=0.01  \citep{sweigart90}\\

       &       & &AGB-Mira & [$-5.0$,$-7.1$]&             &          &Observed  bulge stars in  \citet{alard01} \\
       &       & &AGB-Mira &             &  $<3500$     & M4-M9    & Observed range in the Bulge \citep{blanco84}\\
0.85$^c$  &  11.8$^c$ & &AGB-Mira &             &  $<3500$     & M4-M5    & Observed range in old 47 Tuc \citep{glass73,skiff16}\\
       &       & &AGB-SR   & [$-2.5$,$-5.0$] &              &          & Observed. Bulge stars in  \citet{alard01} \\

1      & 11250    & 12&AGB &$[-3.61,-4.03]$& &&\Mbol\ during E-AGB and TP-AGB by \citet{wood93}\\
2      &  1236    & 9 &AGB &$[-3.78,-4.90]$& &&\Mbol\ during E-AGB and TP-AGB by \citet{wood93}\\
3.5    &   230    & 3 &AGB &$[-5.17,-5.65]$& &&\Mbol\ during E-AGB and TP-AGB by \citet{wood93}\\
5      &    95    & 1.4&AGB &$[-5.91,-6.22]$& &&\Mbol\ during E-AGB and TP-AGB by \citet{wood93}\\

7      &       & &S-AGB     &[$-6.86$] &                 & & minimum \Mbol$^a$  \citet{doherty15}\\
8      &       & &S-AGB     &[$-7.20$] &                 & & minimum \Mbol$^a$  \citet{doherty15}\\
9      &       & &S-AGB     &[$-7.60$] &                 & & minimum \Mbol$^a$  \citet{doherty15} \\
9.8    &       & &S-AGB     &[$-7.86$] &                 & & minimum \Mbol$^a$  \citet{doherty15}\\
3      & 417   &    & S-AGB & [$-0.3$,$-1.7$]   & 4850 - 4300 & $>$K0      & Rot. tracks$^b$  by \citet[][]{ekstroem12}\\
5      & 111   &    & S-AGB & [$-2.3$,$-4.4$]   & 4600 - 3800 & $>$K0 - M0 & Rot. tracks$^b$ by \citet[][]{ekstroem12}\\
7      &  52   &    & S-AGB & [$-3.5$,$-5.9$]   & 4400 - 3550 & $>$K0 - M3.5 & Rot. tracks$^b$ by \citet[][]{ekstroem12}\\
9      &  32   & 3.7  &RSG  & [$-4.5$,$-6.8$]   & 4200 - 3500  & K0 - M4.5  & Rot. tracks by \citet[][]{ekstroem12}\\
12     &  20   & 2.0  &RSG  & [$-6.0$,$-7.4$]   & 3900 - 3550  & K4 - M3.5  & Rot. tracks \citet[][]{ekstroem12}\\
15     & 12.5  & 1.0  &RSG  & [$-7.3$,$-7.9$]   & 3750-  3600  & M1 - M2    & Rot. tracks \citet[][]{ekstroem12}\\
20     &  9.9  &      &RSG  & [$-8.2]$          & 3774         & M0.5       & Rot. tracks \citet[][]{ekstroem12} \\
25     &  8.0  &      &RSG  & [$-8.79$]         & 3836         & K5         & Rot. tracks \citet[][]{ekstroem12}\\
       &       &      &RSG  & [$-3.63$,$-10.36$]&              &            & Observed range by \citet{levesque05} \\
\hline
\end{tabular}
\begin{list}{}
\item
$(^a)$ during the interpulse phase.\\
$(^b)$ evolved up the early asymptotic giant branch.\\
$(^c)$ age of 47 Tuc \citep{brogaard17}. \\

\end{list}
\end{table*}

\begin{figure*}
\begin{center}
\resizebox{0.49\hsize}{!}{\includegraphics[angle=0]{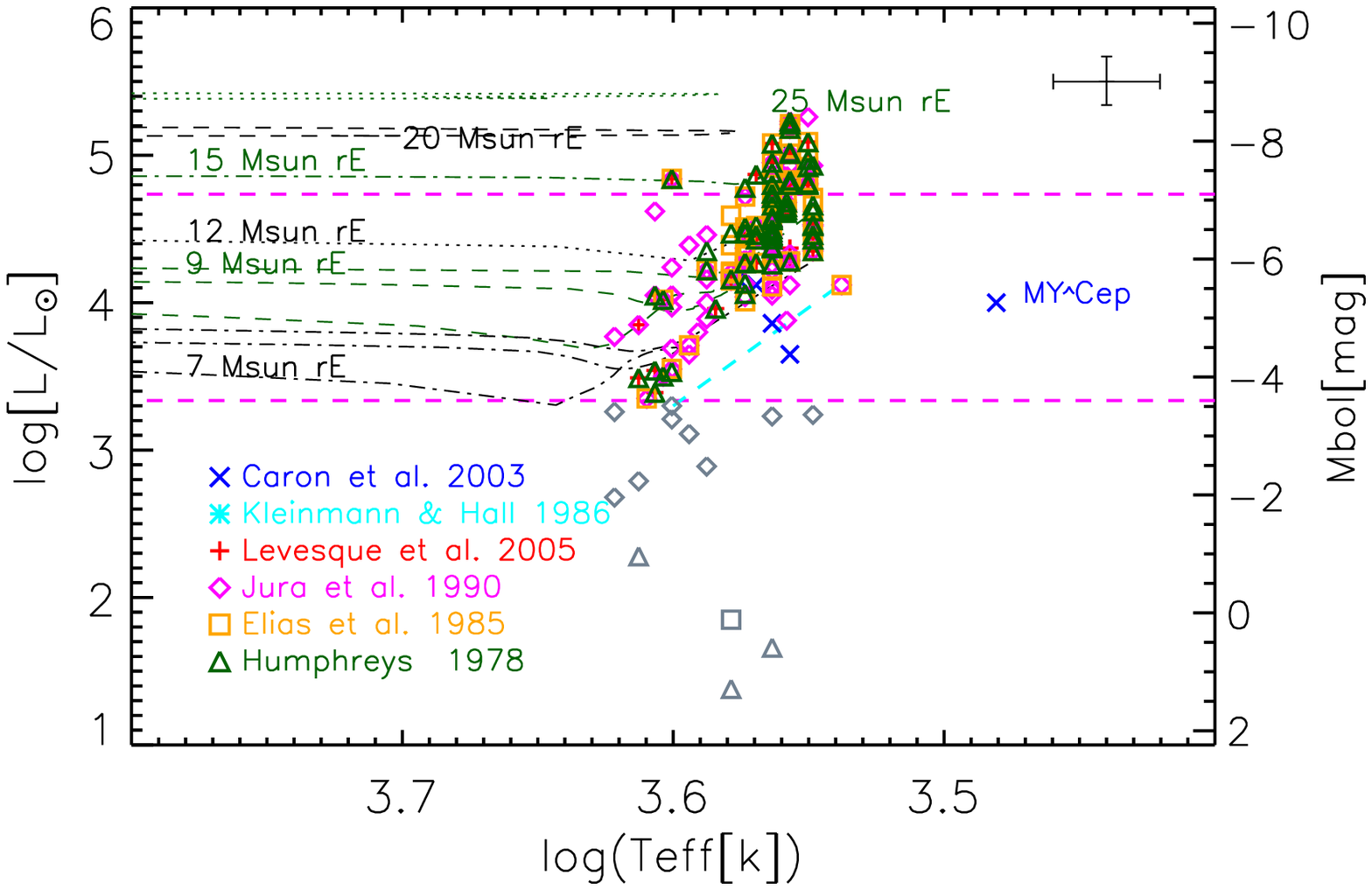}}
\resizebox{0.49\hsize}{!}{\includegraphics[angle=0]{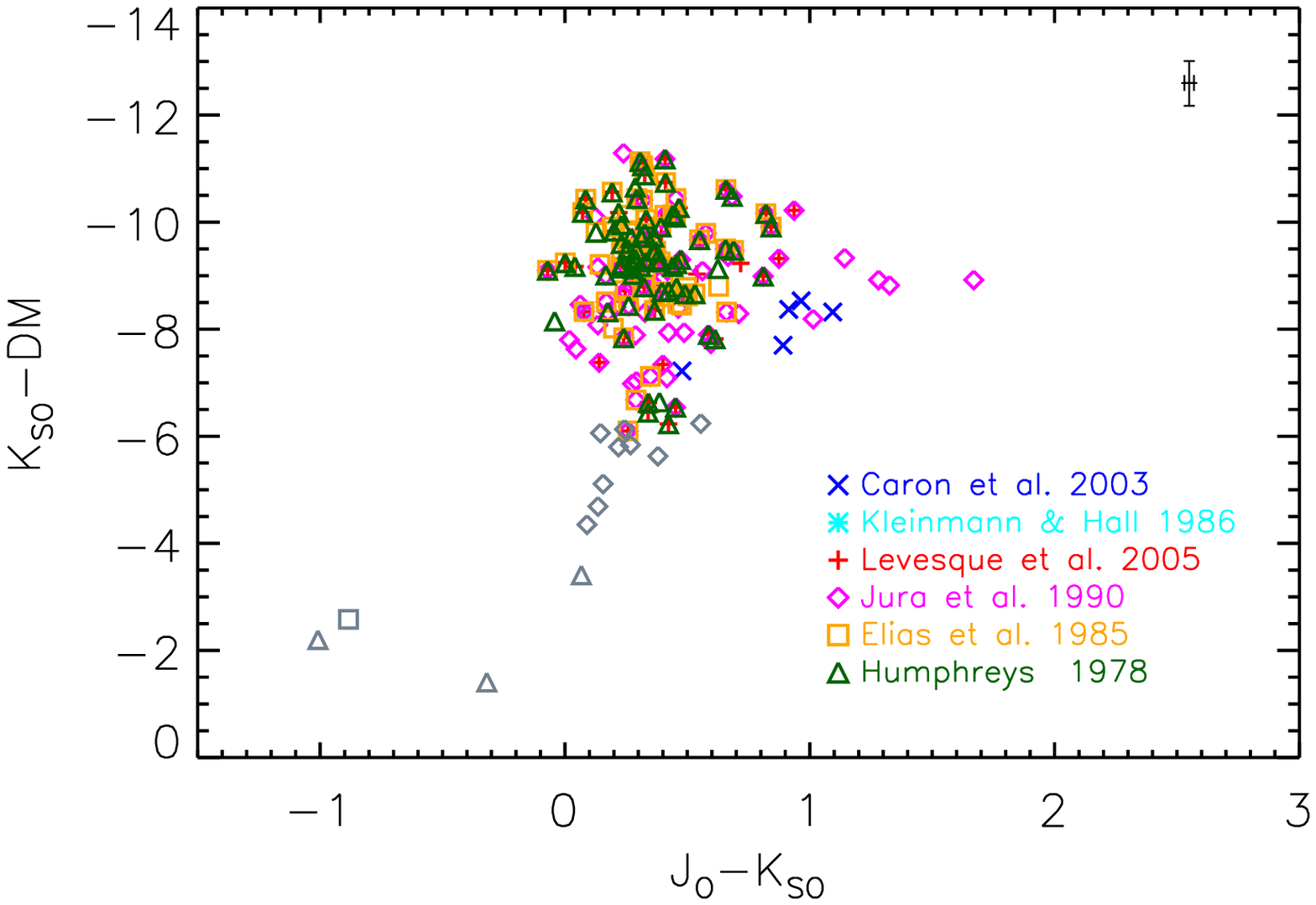}}
\resizebox{0.49\hsize}{!}{\includegraphics[angle=0]{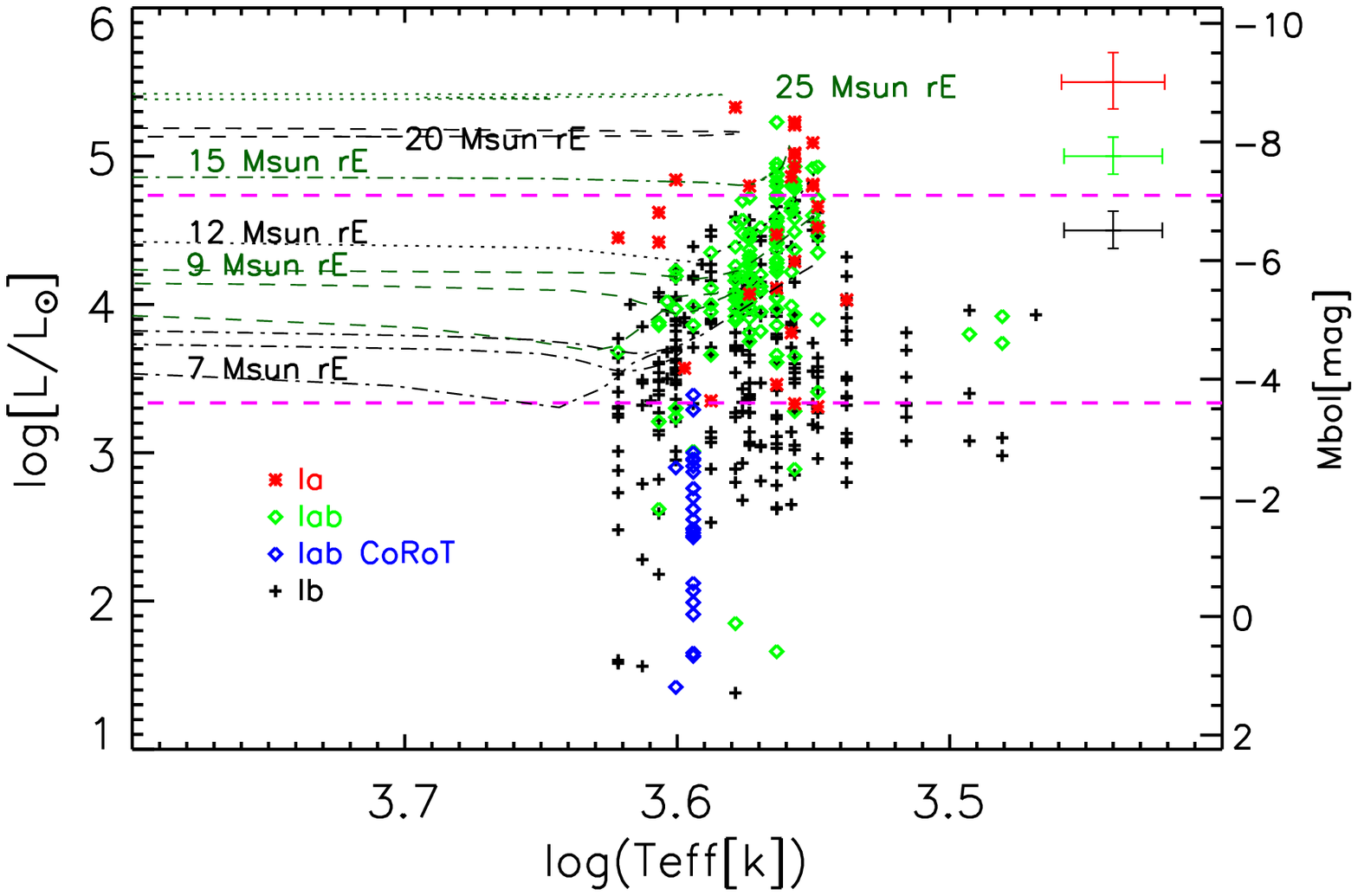}}
\resizebox{0.49\hsize}{!}{\includegraphics[angle=0]{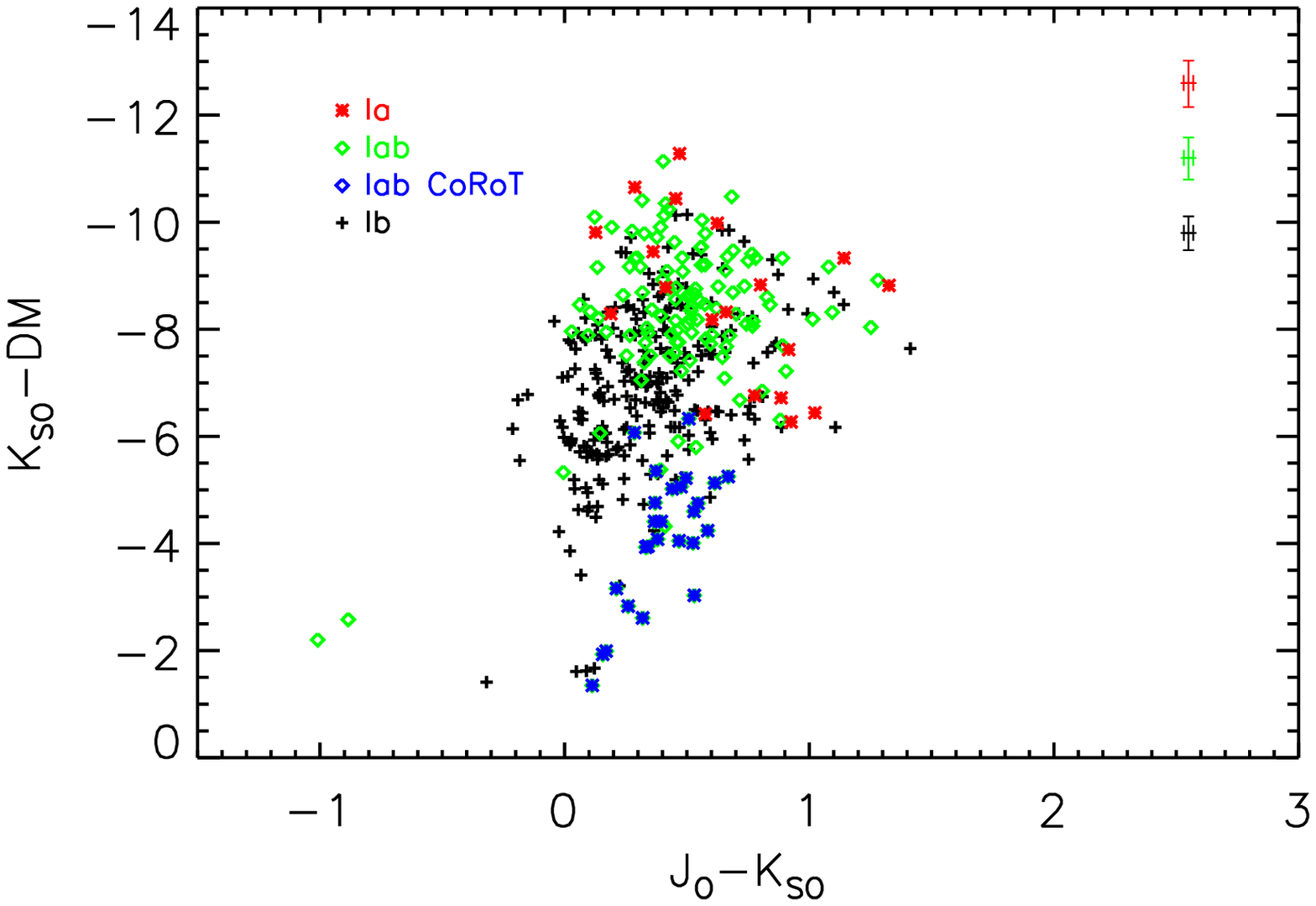}}

\caption{ \label{compare_ref} {\it Left-upper Panel:} 
Luminosities versus \Teff\ values of reference RSGs (from class Ia to Ib), i.e. of the subsample of stars 
in Table \ref{table.gaia} with given class I in the catalogs of \citet[][blue crosses]{caron03}, 
\citet[][cyan asterisck]{kleinmann86}, 
\citet[][red pluses]{levesque05}, \citep[][magenta diamonds]{jura90}, 
\citet[][orange squares]{elias85}, and \citet[][green triangles]{humphreys78}.
An average error bar is drawn on the right-upper corner.
The two magenta long-dashed horizontal lines marks \Mbol=$-3.6$ mag (tip of the red giant branch),
and $-7.1$ mag (AGB limit).
The long-dashed cyan line marks the Eq. (1);  RSGs appear brighter and bluer than that locus  (see text).
Stellar tracks from models at solar metallicity and including rotation are from \citet{ekstroem12};
from the bottom to the top: the black dotted-dashed curve marks a stellar track of a 7 \Msun\ star; 
the green long-dashed curve marks a 9 \Msun\ track; the black dotted curve a 12 \Msun\ track;
the green dotted-dashed curve shows a 15 \Msun\ track; the black long-dashed the track of 
a 20 \Msun\ track, and the top green dotted line that of a 25 \Msun.
A few objects (in grey) remain fainter than the red giant tip (see text).
{\it Right-upper Panel:} Absolute and dereddened \Ks\ magnitudes versus de-reddened $J-$\Ks\ colors.
Data points are as described in the left panel. 
{\it Right-lower Panel:} Luminosities versus \Teff\ values of  stars in Table \ref{table.gaia} with adopted class Ia, Iab, and Ib
(stars detected by CoRoT and listed in Table \ref{table.gaia} as class Iab  should be regarded separately).
{\it Left-lower Panel:} Absolute and dereddened \Ks\ magnitudes versus de-reddened $J-$\Ks\ colors of  stars in Table 
\ref{table.gaia} with adopted class Ia, Iab, and Ib.
}
\end{center}
\end{figure*}

\subsubsection{Reference RSGs}
We consider as reference RSGs those stars included in the catalogs of
\citet[][]{kleinmann86},
\citet[][]{levesque05}, \citet[][]{caron03}, \citet[][]{jura90}, \citet[][]{elias85}, 
and \citet[][]{humphreys78}.
These sources are expected to be RSGs, because they are located in the direction of OB associations. 
In the upper-left panel of Fig.\ \ref{compare_ref}, we show their luminosities, $\log(L/$\Lsun), versus
\Teff\ (theoretical plane); in the lower panel, we show their absolute and dereddened \Ks, $K_{\rm so}-$DM versus $J_{\rm o}-K_{\rm so}$
(observational plane); DM is the distance moduli. By comparison with the stellar tracks, we estimated initial masses from
about 7 to 25 \Msun\ \citep{ekstroem12}. Among them, the brighest star appears to be SW Cep with
\Mbol= $-8.42$ mag. MY-Cep is the only M7.5 I included in the sample. A few stars were discarded
as reference RSGs, because they appeared too faint for luminosity class I (\Mbol\ $ > -3.6$ mag, as
shown in Fig.\ \ref{compare_ref}); those stars are  IRC+40105,
6~Aur, 1~Pup, sigOph, IRC~+00328, 33~Sgr, 12~Peg, BD+47~3584, 56~Peg \citep{jura90}, CD-57~3502
\citep{elias85}, CPD-59~4549, HD~142686, and HD~150675 \citep{humphreys78}.

\begin{figure*}
\begin{center}
\resizebox{0.49\hsize}{!}{\includegraphics[angle=0]{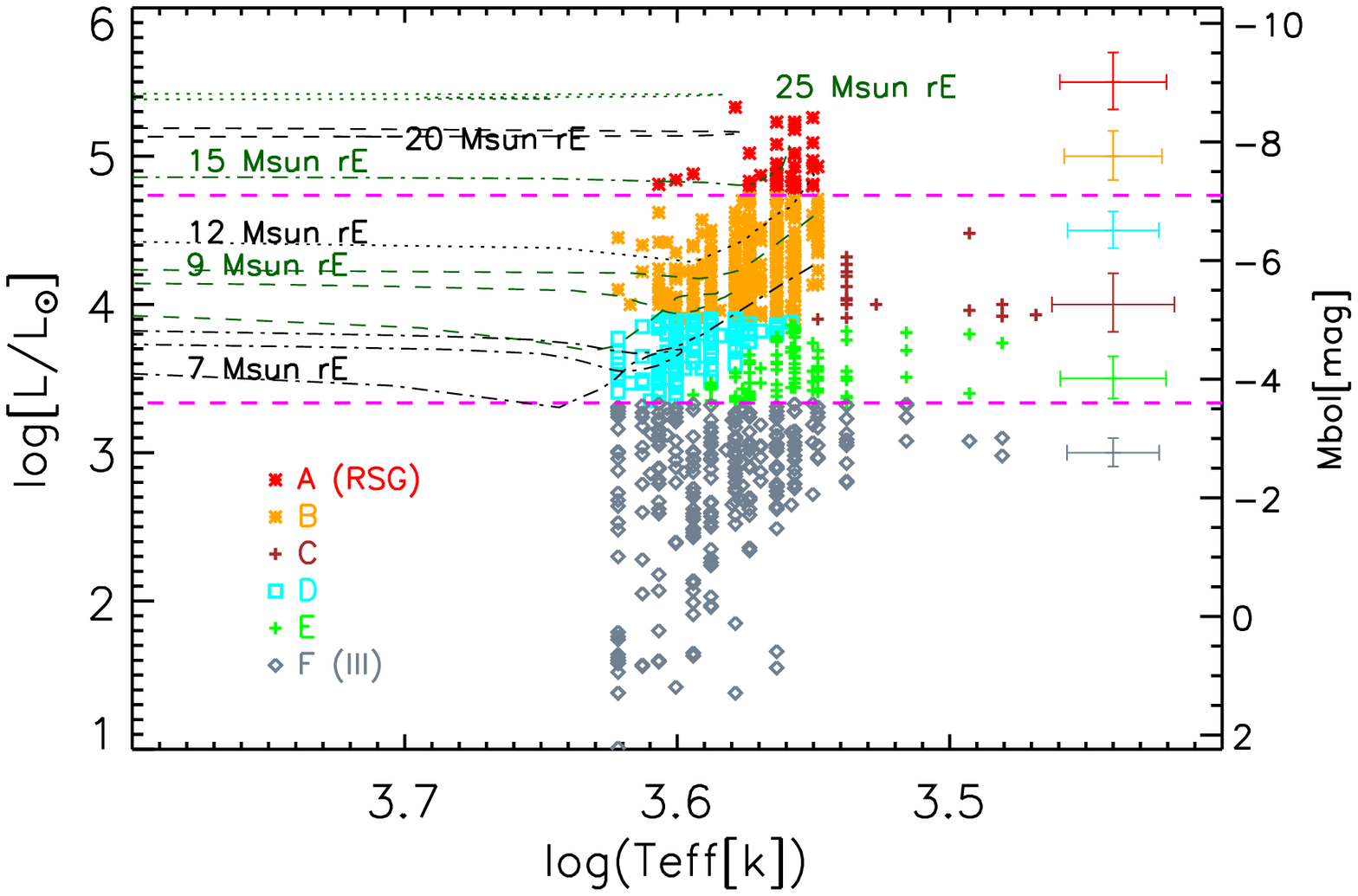}}
\resizebox{0.49\hsize}{!}{\includegraphics[angle=0]{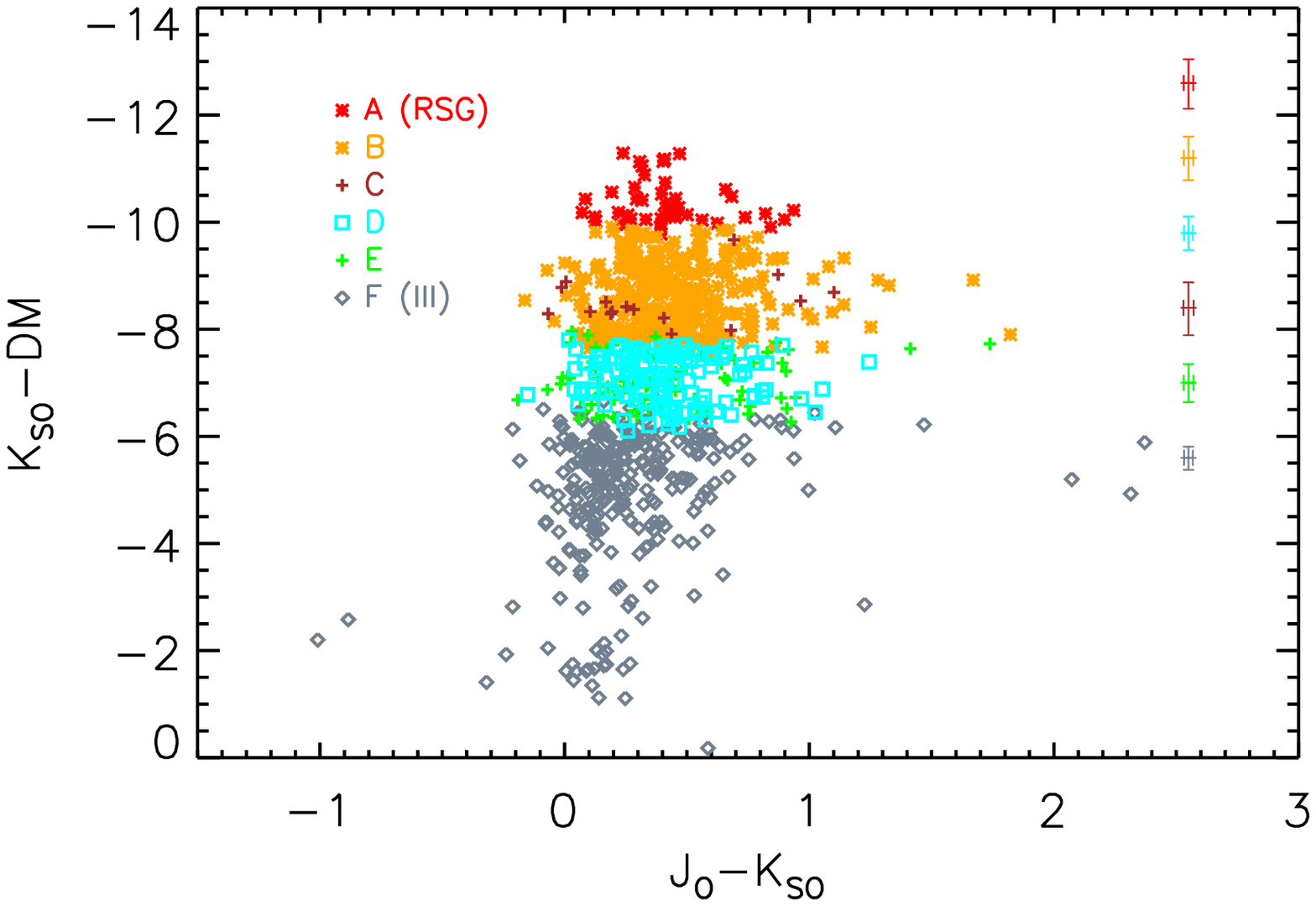}}
\caption{ \label{compare}  {\it Left Panel:} 
Luminosities versus \Teff\ values of stars in Table \ref{table.gaia}
with $\varpi/{\sigma_\varpi}({\rm ext}) > 4$ and $\text{RUWE} < 2.7$. 
Red asterisks mark highly-probable RSGs with \Mbol $< -7.1$ mag   (Area=A).
Orange asterisks mark sources with \Mbol $< -5.0$ mag and types $< M4$ (Area=B).
Cyan squares mark sources  with $-3.6>$\Mbol$>-5.0$ mag and bluer that Eq. (1) (Area=D).
Brown pluses (\Mbol$< -5.0$ mag) and green pluses ($-3.6>$\Mbol$>-5.0$ mag)
indicate the Areas C and E.
 Gray diamonds indicate giants, i.e., stars
fainter than \Mbol$\approx -3.6$ mag (tip of the red giant branch,  Area=F).
The two magenta long-dashed horizontal lines mark \Mbol=$-3.6$ mag (tip), 
mag, and \Mbol=$-7.1$ mag (AGB limit).
For comparison, we add some rotating stellar tracks with solar metallicity by \citet{ekstroem12}. 
From the bottom to the top: the black dotted-dashed curve marks a stellar track of a 7 \Msun\ star; 
the green long-dashed curve marks a 9 \Msun\ track; the black dotted curve a 12 \Msun\ track;
the green dotted-dashed curve shows a 15 \Msun\ track; the black long-dashed the track of a 20 \Msun\ track, 
and the top green dotted line that of a 25 \Msun.
{\it Right Panel:} Absolute and dereddened \Ks\ magnitudes versus de-reddened $J-$\Ks\ colors. 
Data points are as described in the left panel. }

\end{center}
\end{figure*}

\subsubsection{Hertzsprung-Russell diagram}
All reference   RSGs, but  MY~Cep, appear
located along the ascending stellar tracks in Fig. \ref{compare_ref}.
They are located to the left of the following equation 
(which is roughly parallel to the  ascending
parts of the tracks at the low \Teff\ end): 
\begin{equation}
  \log(L/\Lsun) = 51.3-13.33\times \log(\Teff)\,,
\end{equation}
where $\log(\Teff)$ ranges from 3.54 to 3.6 \citep[i.e., from M4 to K1,][]{levesque05}.

The temporal evolution of an AGB star is characterised by large excursion  in the
\Mbol\ versus \Teff\ diagram. During the thermal pulses the luminosity increases and 
\Teff\ decreases. For example, a star 3 \Msun\ may reach \Mbol=$\approx -2$ mag during 
the early-AGB phase and \Mbol=$\approx -5$ mag during thermal pulses (e.g., \citet{wood93}).

In Fig. \ref{compare}, we show the luminosities of stars in Table \ref{table.properties}, 
and we verify their positions on the \Mbol\ versus \Teff\ diagram by using the described observational benchmarks and
the features appearing in Fig.\ \ref{compare_ref}:
\begin{itemize}

\item[$A$]  Area A contains late-type stars with \Mbol $ \la -7.1$ mag, 
those  are expected to be mostly RSGs.

\item[$B$] Area B  contains stars with $-5.0>$\Mbol$>-7.1$ mag and earlier than an M4. 
This area is rich in stars with masses larger than 7 \Msun. 

\item[$C$] Area C  contains late-type stars with $-5.0>$\Mbol$>-7.1$ mag and
later than an M4.  This area is expected to be dominated by AGBs (4-9 \Msun).

\item[$D$] Area D  contains late-type stars with $-3.6>$\Mbol$>-5.0$ mag and
bluer than Eq. (1). This area  contain  AGBs 
of intermediate masses and some faint K-type 9 \Msun\ stars 
at the onset of their cold phase (\Mbol = $-4.5$ mag).

\item[$E$] Area E  contains late-type stars with $-3.6>$\Mbol$>-5.0$ mag and
redder than Eq. (1). This area is expected to be dominated by old and more 
abundant   AGBs (2-3 \Msun).

\item[$F$] Area F  contains late-type stars with \Mbol $> -3.6$ mag.
Those stars are fainter than the tip of the red giant branch.
    
\end{itemize}

\begin{figure*}
\begin{center}
\resizebox{0.49\hsize}{!}{\includegraphics[angle=0]{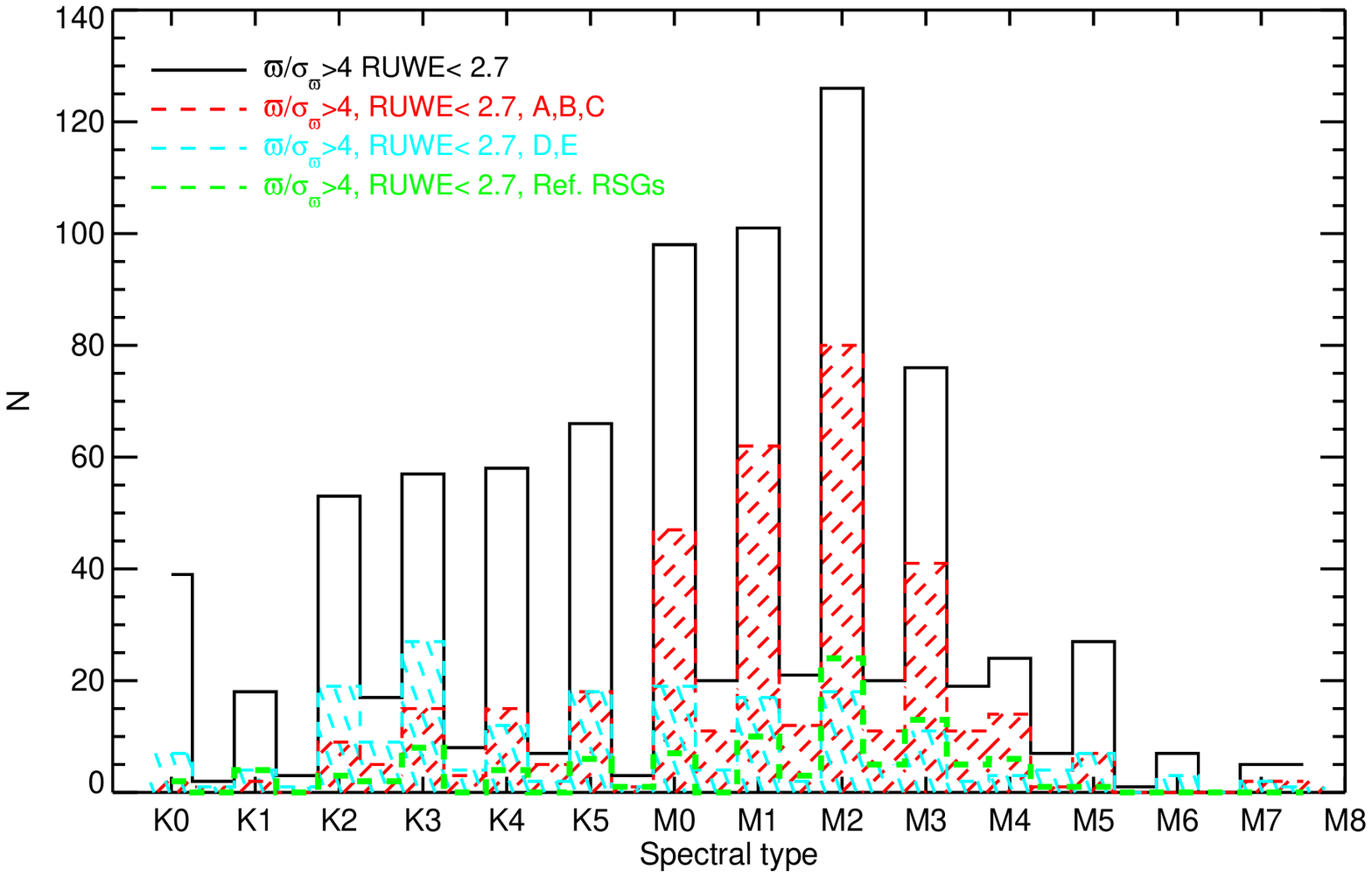}}
\resizebox{0.49\hsize}{!}{\includegraphics[angle=0]{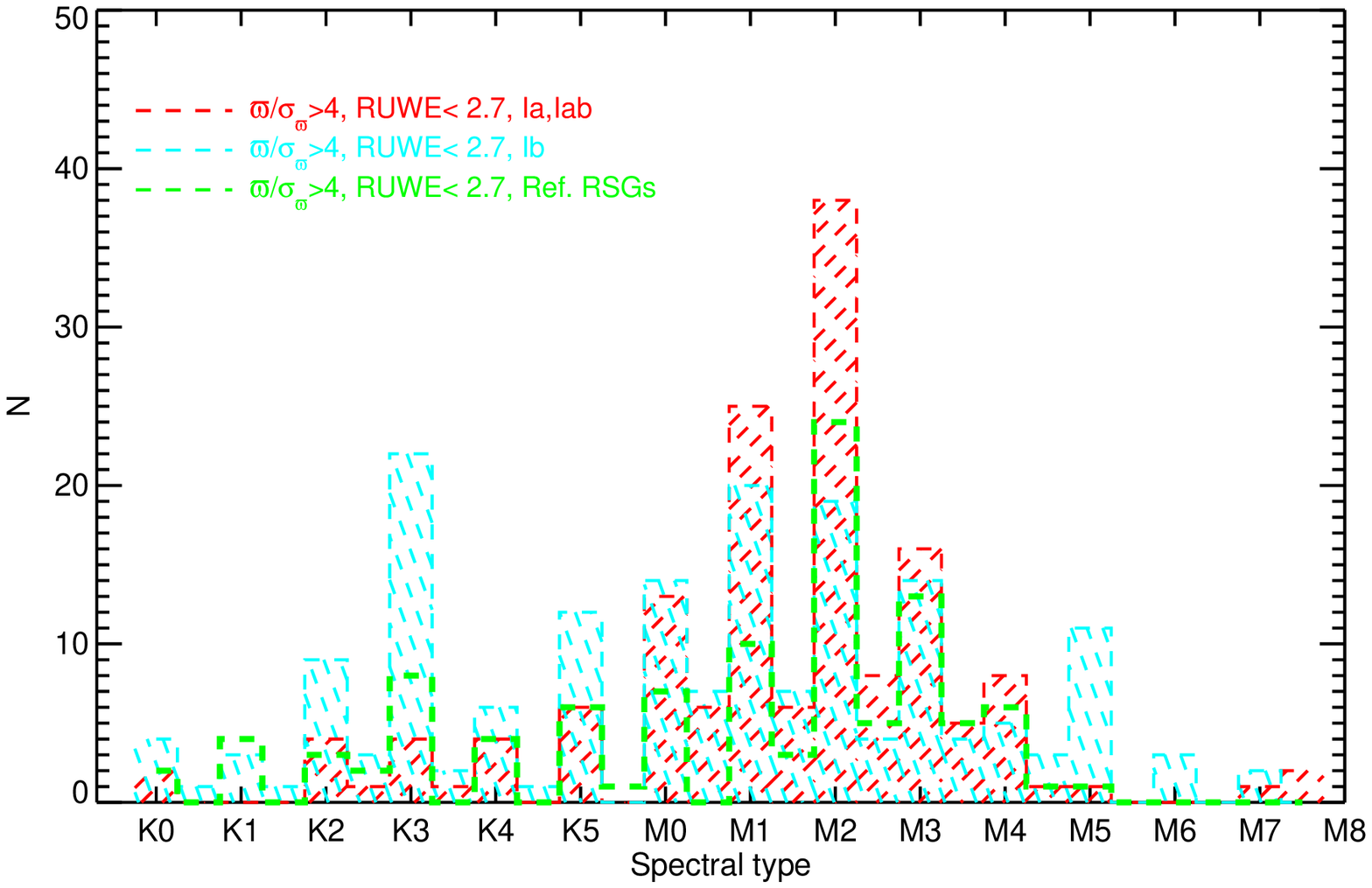}}
\caption{ \label{sphistogram} {\bf Right panel:}  
In black the histogram of the spectral types of sources with 
good distances ($\varpi/{\sigma_\varpi}({\rm ext})>4$ and $\text{RUWE} < 2.7$);
in red that of sources with \Mbol$<-5.0$ mag, i.e located in  Area A and B, 
or  in Area C but reported with class I in all previous literature; 
in cyan the histogram of sources with \Mbol$>-5.0$ mag, located in 
 Area D, or in Area E but reported with class I in all previous literature;
in green the histogram of  reference RSGs.
{\bf Right panel:} in red sources of adopted class Ia,  Iab and with good distances and \Mbol$<-3.6$ mag;
in cyan the histogram of sources of adopted class Ib with good distances and \Mbol$<-3.6$ mag;
in green the histogram of  reference RSGs.
}
\end{center}
\end{figure*}

 In Fig. \ref{compare}, in the theoretical \Mbol\
versus \Teff\ diagram as well as in the observational $K_{\rm so}-$DM versus $J_o-K_{so}$ diagram, we mark the
areas  defined above with different colors. These luminosity areas are also added
in Table \ref{table.properties}. In Fig. \ref{sphistogram}, we show an histogram of the 
spectral types of the 889 sources with $\varpi/\sigma_\varpi >4$ and $\text{RUWE} < 2.7$. 

Reference RSGs appear to be made by stars with class Ia and Iab (35\%),
as well as of stars with class Ib (33\%).
In Figs. \ref{compare_ref} and \ref{sphistogram}, the distribution of 
reference RSGs appears similar
to that of stars Ia and Iab, with  stars falling mostly in the Area A and B;
but it's different than that of class Ib stars, 
which are  sparsely distributed over  the Area A,B,C,E, and F.

From Table \ref{table.gaia}, about 43 sources (5\%) are found to 
be located in the  Area A (\Mbol $\la -7.1$ mag). 
Among them there are two stars, HD 99619 and HD 105563 A, with 
previous uncertain class.
312 sources (35\%) are located in Area B and are likely more massive than 7 \Msun.
About 30\% of the sample is made of stars fainter than the tip of the red giant branch (Area F).

A large number of RSGs detected at infrared wavelength (about 300) was included 
in the presented compilation; however for most of those stars parallaxes are not available in DR2 
(Table \ref{statistic} shows only 16 stars from infrared catalogs), \citep[for
example,][]{davies08,davies07,liermann09,clark09,negueruela10,
negueruela11,negueruela12,messineo17}.

\subsection{Gaia variables}

We searched our sample for the presence of Gaia variables and found that only 137 stars of the
initial 1342 source with Gaia data were flagged as variables \citep{holl18},  and 90 out of the 889
with good parallaxes (about 10\%). The spectral types of all 90 but one variables range from K5 to M7,
and 83 of them are automatically classified by the Gaia pipeline as long period variables, LPVs,
including Mira and semiregular (SR) stars. 
Their average variation in $G$-band is 0.51 mag with a dispersion around the mean of 0.38 mag,
including two stars with variations above 2.5 mag (0.1\%), which are in Area C and E.
There are 65 (out of 90) variables in Area A and B; their variations in $G$-band range from 
0.2 mag to 0.8 mag, with a mean variation of 0.41 mag and dispersion around the mean 
of 0.14 mag. Similar values are found with the 9 variables of class Ib 
(a mean of 0.46 mag and a $\sigma=0.33$ mag). There are 9 variables fainter than 
\Mbol $>-3.6$ mag  (Area F), with 7 of them later than M5.
Their mean variation is 0.63 mag and $\sigma= 0.45$ mag.

An analysis of the $G$-band light curves will be presented elsewhere.

\begin{table*}
\caption{\label{statistic} Numbers of collected stars per luminosity classes.}
\begin{tabular}{l|rrr|rrrrrrrrr}
\hline
\hline
Sample     &N(sp)  & N(Ks) & N(plx)   & \multicolumn{5}{c}{\rm N(Ks+plx)}  \\
           &     &       &           &   N$_{\rm A}$ & N$_{\rm B}$   & N$_{\rm D}$   & N$_{\rm CE}$  & N$_{\rm F}$ &  Nnew(I) \\
           &     &       &           &               & blue          & blue          & red           & (III) &                                   \\
\hline
All$^a$           &1406&1406 & 889   & 43 & 322 & 134&110         & 280        &     35  \\
Ref. opt stars$^b$& 170& 170 & 135   & 26 &  69 &  21&    2&  17&     0  \\
Ref. IR stars $^c$& 312& 312 &16     &  0 &   1 &  3 & 0  &  12 &      1  \\
Nsp(Ia)           & 57 &  57 & 28    & 12 &   9 & 1 &4 &2 &0\\
Nsp(Iab)          & 243& 243&161& 16& 90&11& 9& 35 &0\\
Nsp(Ib)           & 300& 300&259&  2& 76&52&48& 81 &0\\
Nsp(any I)        &1013&1013&620& 41&253&86&82&158 &0\\
Nsp(I-II)         & 166& 166&113&  0& 36&24&14& 39 &0\\
\hline
\end{tabular}
\begin{list}{}
\item {\bf Notes.}
N(sp) = number of stars with known available spectral types.~ 
N(Ks) = number of stars with available near-infrared measurements.~
N(plx)= number of stars with $\varpi/{\sigma_\varpi}({\rm ext}) >4$ and $\text{RUWE} < 2.7$.~
N$_{\rm A}$= number of stars located in  Area A.~ 
N$_{\rm B}$= number of stars located in  Area B.~ 
N$_{\rm D}$= number of stars located in  Area D.~ 
N$_{\rm CE}$= number of stars located in  Area C or E.~ 
N$_{\rm F}$ = number of stars  in  Area F.~ 
Nnew(I)=number of stars without  adopted classes  and to which we assign  Area A or B.~

Nsp(Ia)  =  number of stars with  luminosity classes Ia.~      
Nsp(Iab) =  number of stars with  luminosity classes Iab.~   
Nsp(Ib)  =  number of stars with  luminosity classes Ib.~ 
Nsp(any I)= number of stars with  luminosity classes (I,Ia,Iab,Ib).~   
Nsp(I-II) = number of stars with  luminosity classes (I-II).~    
($a$) All stars in Table \ref{table.gaia}.~
($b$) Example of optically visible RSGs taken from 
\citet[][]{caron03}, 
\citet[][]{levesque05}, \citet[][]{jura90},  \citet{kleinmann86},
\citet[][]{elias85}, and \citet[][]{humphreys78}.~
($c$) Example of optically obscured sources taken 
from \citet{messineo17}, \citet{clark09},  \citet{davies08},  \citet{davies07},
 \citet{negueruela12}, \citet{negueruela11}, \citet{negueruela10}, and \citet{liermann09}. 
\end{list}
\end{table*}

\subsection{Average magnitudes per spectral type.}

In Table \ref{table.magbin} we present average magnitudes per spectral type of stars of class I
and with \Mbol $< -5.0$ mag, and of stars with $-3.6<$\Mbol$<-5.0$ mag.
This table is useful for Galactic star counts \citep[e.g.][]{wainscoat92}. In Table 2 of
\citet{just15} infrared luminosities of Hipparcos stars per classes are also provided; for example,
their K-M2 I-II stars have \Mk= $-9$ mag. For  stars with spectral types K-M2 I and \Mbol$<-5.0$ mag,
our Table \ref{table.magbin} provides an average \Mk= $-8.40$ mag with $\sigma$=0.39 mag.

Additionally, in Tables \ref{table.magbinIaIab} and \ref{table.magbinrefRSG} 
we present average magnitudes per spectral type of stars of classes Ia and Iab
and of stars in the reference RSG sample. 

\begin{figure*}
\resizebox{0.7\hsize}{!}{\includegraphics[angle=0]{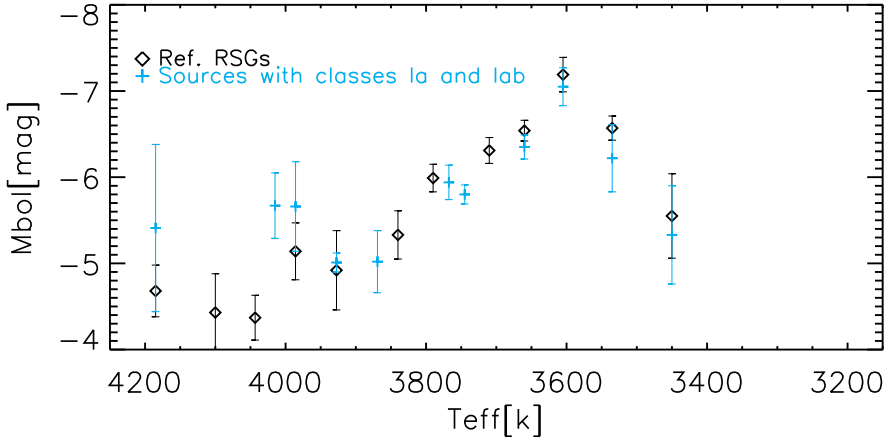}}
  \caption{\label{fig:mbolbin}
   Average \Mbol\ versus \Teff. Cyan crosses show the values for class Ia and Iab stars.
   Black diamonds indicate the values for the reference RSGs. }
\end{figure*}

In Fig. \ref{fig:mbolbin}, we plot the calculated average magnitudes per spectral 
types of stars with classes Ia and Iab, as well as of stars in the reference RSG sample,
versus the \Teff\ values. \Teff\ were estimated from the spectral types with the 
temperature scale given by \citet{levesque05}.
For stars with \Teff\ from  3650 k to 3950 k, \Mbol\ values seem 
to decrease with decreasing \Teff\ values.

\subsection{Spatial distribution}
\label{spatialdistribution}

\begin{figure*}
\resizebox{0.33\hsize}{!}{\includegraphics[angle=0]{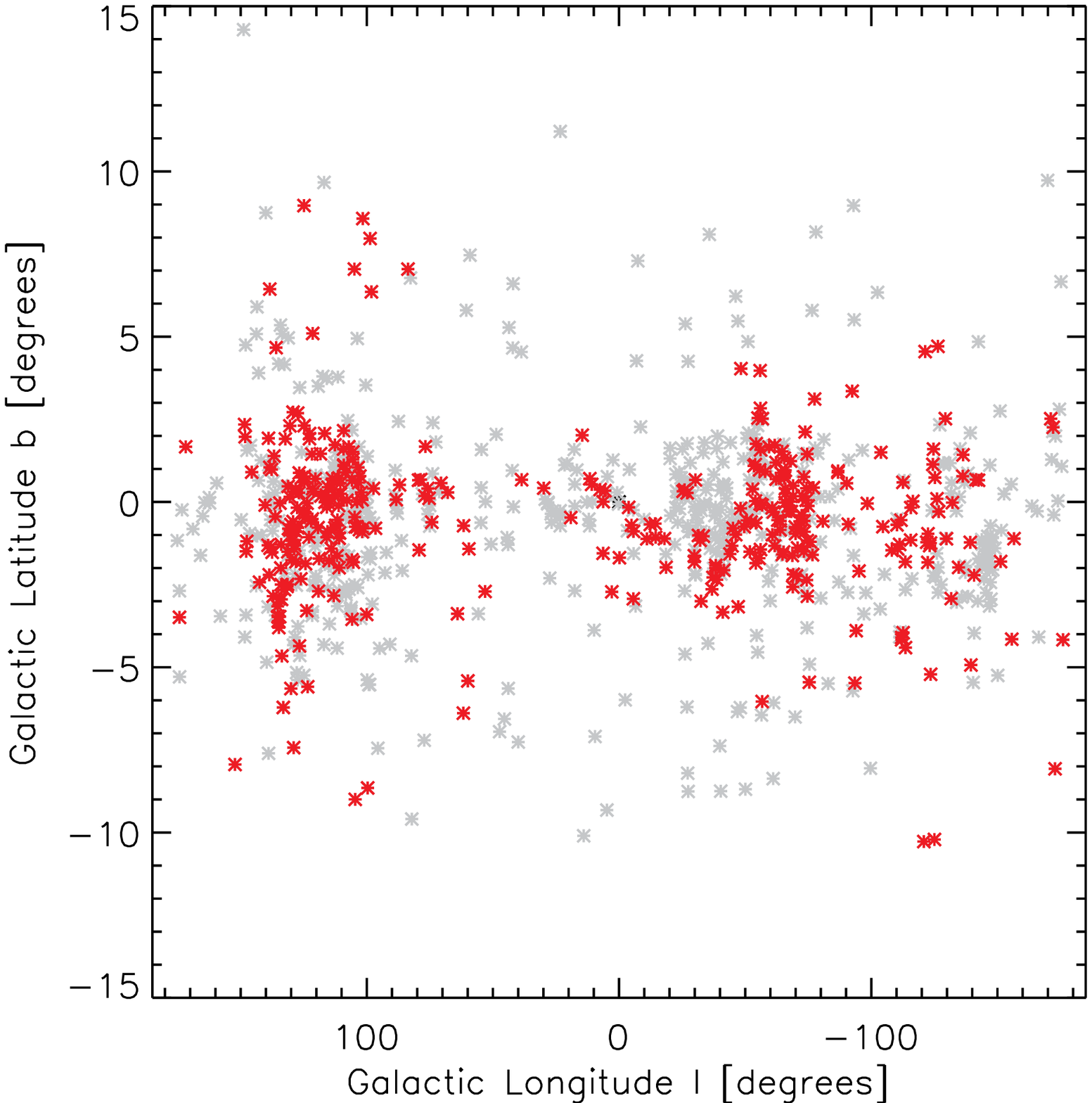}}
\resizebox{0.33\hsize}{!}{\includegraphics[angle=0]{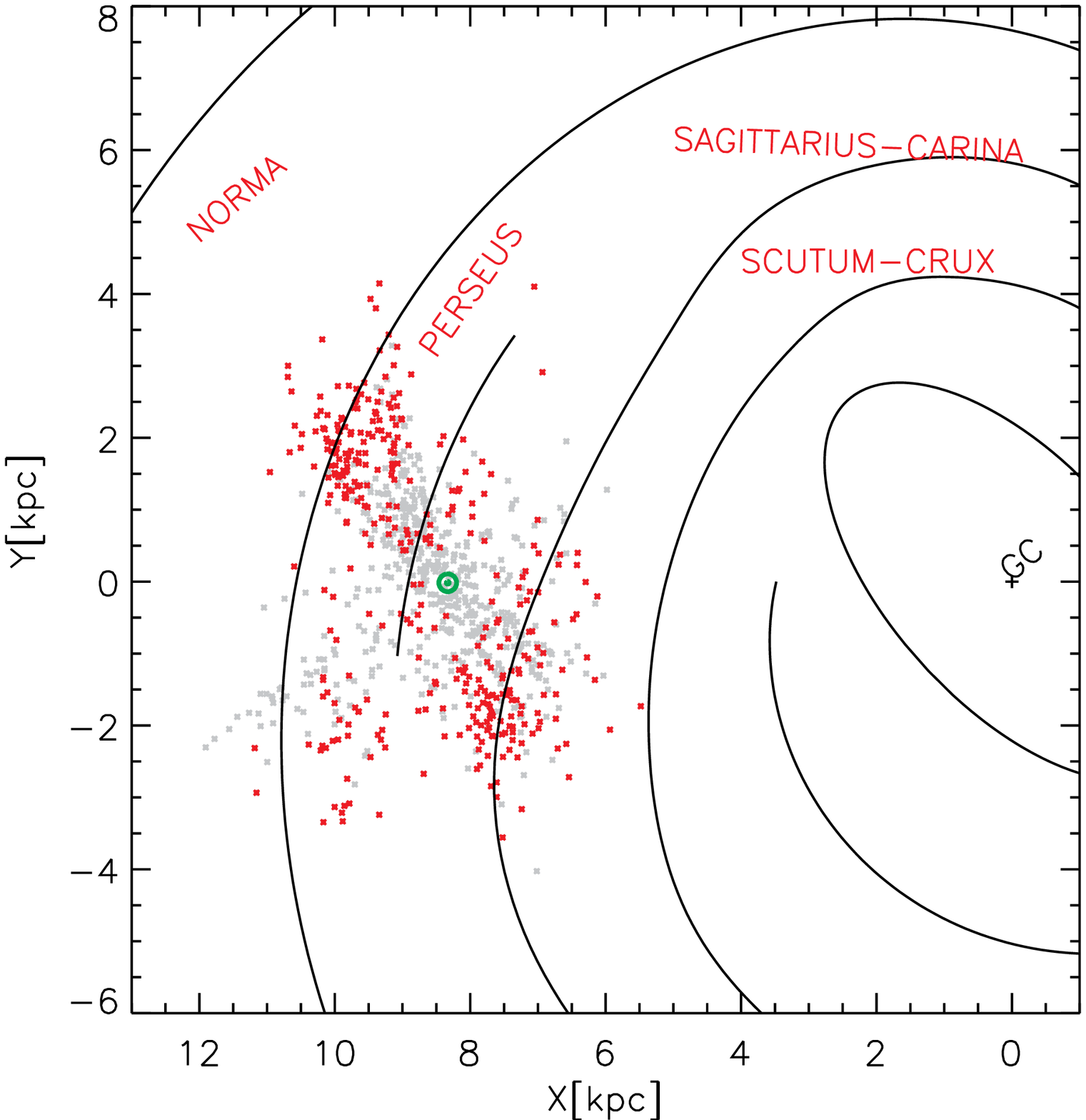}}
\resizebox{0.33\hsize}{!}{\includegraphics[angle=0]{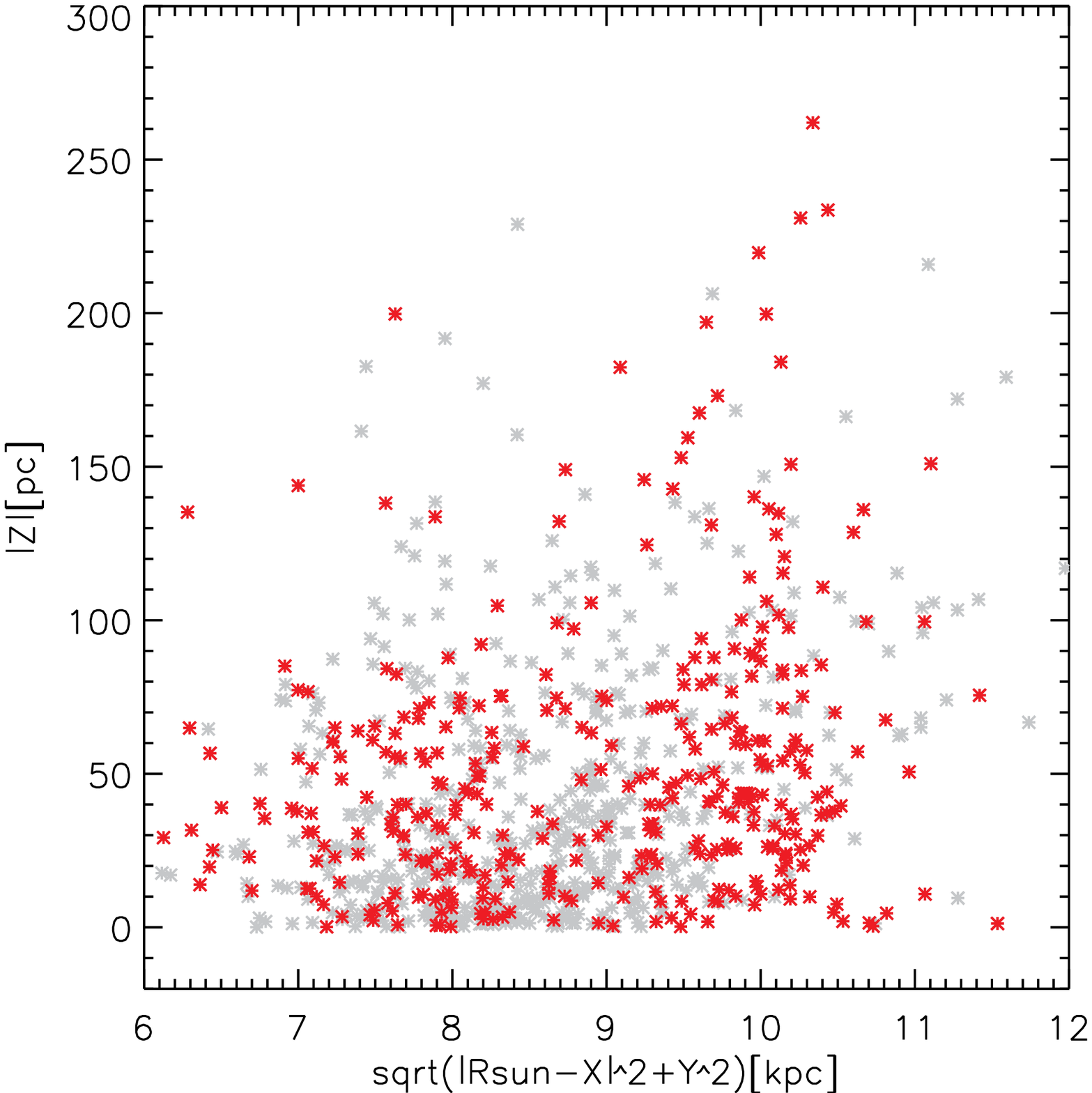}}
\resizebox{0.33\hsize}{!}{\includegraphics[angle=0]{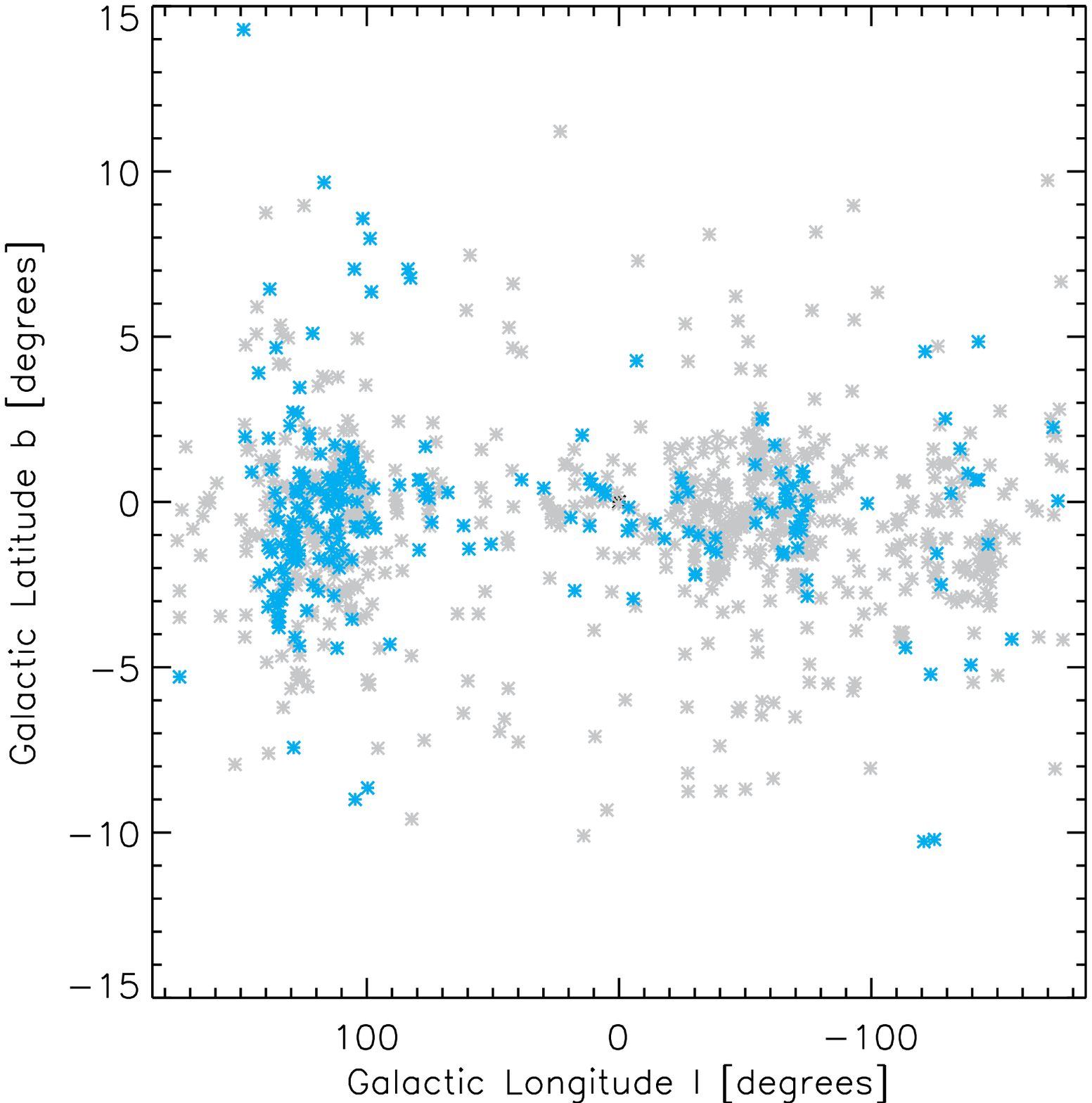}}
\resizebox{0.33\hsize}{!}{\includegraphics[angle=0]{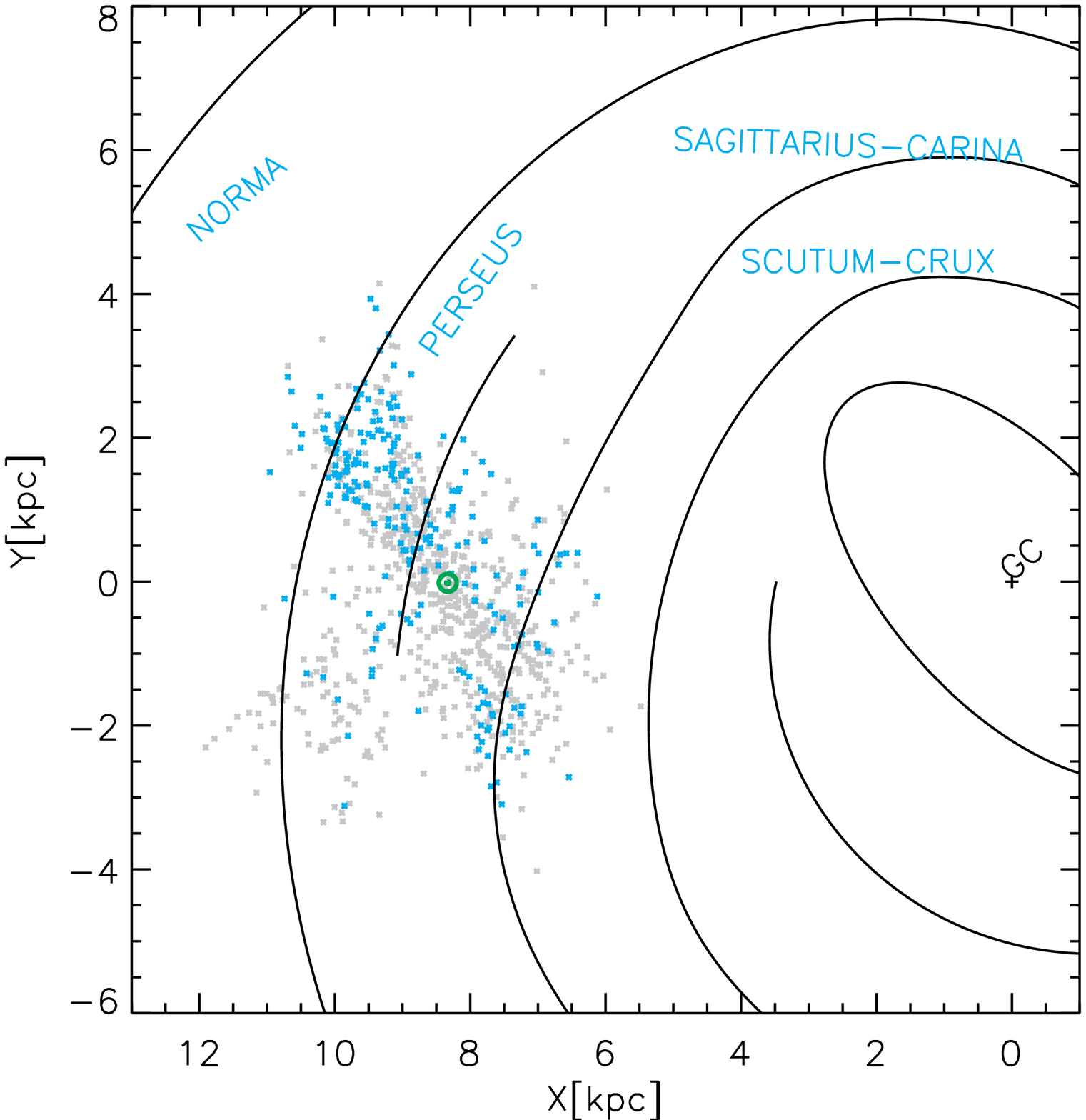}}
\resizebox{0.33\hsize}{!}{\includegraphics[angle=0]{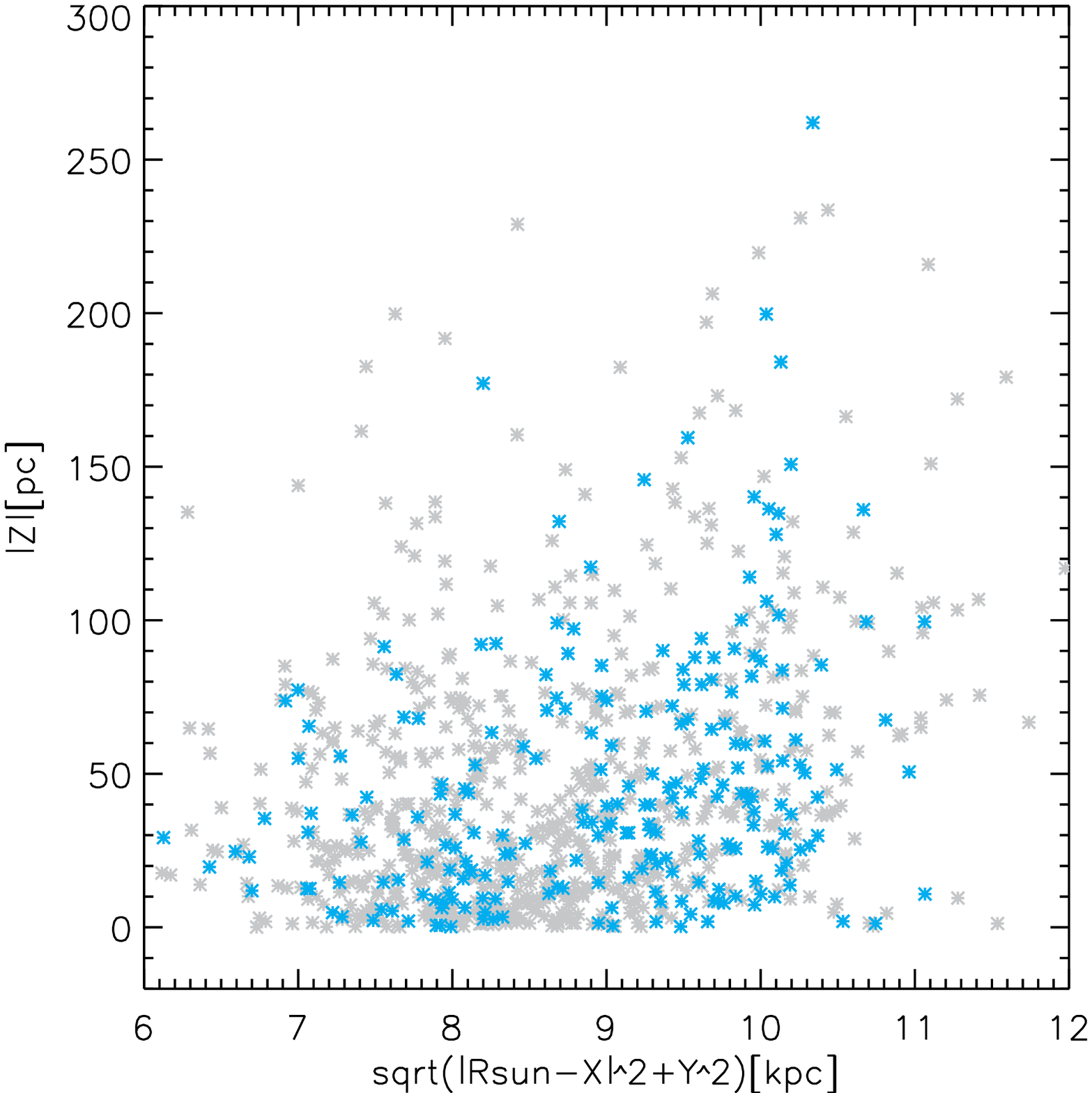}}
\caption{ \label{lb}  {\it Left-upper panel:} Latitudes versus longitudes of the 
bright late-type stars in Table \ref{table.gaia}. 
Candidate RSGs with \Mbol $<-5.0$ mag  (Area A and B)  
and $\varpi/{\sigma_\varpi}({\rm ext})> 4$ and $\text{RUWE} < 2.7$ are marked in red. 
{\it Central-upper panel:} Galactocentric coordinates $XY$  on the disk of the Milky Way. 
The Sun location (8.5,0) is marked in green, 
while the Galactic Centre (GC) marked with a black cross is at (0,0). 
The  spiral arms are taken from the work of \citet{cordes03}. 
{\it Right-upper panel:} 
Distances from the plane $|Z|$ versus Galactocentric distances.
{\it Lower panels:} As in the uppe panels, but this time  cyan asterisks mark  
bright late-type stars in Table \ref{table.gaia} with class Ia or Iab,  or  reference RSGs (see Fig. \ref{compare_ref}). }
\end{figure*}

The bright cool stars here analyzed span 360$^\circ$ of longitude (Fig. \ref{lb}). By using the
estimates of distances in Table \ref{table.gaia}, we obtained the distribution on the Galactic plane
shown in  Fig. \ref{lb}. 
Late-type stars brighter than \Mbol=$-5.0$ mag ($0.8 \times
10^4$ \Lsun) appear radially more distant from the Sun  than the whole sample, 
with heliocentric distances ranging
from $\approx 200$ to $\approx 4600$ pc. Star eta~Per (K3 Ib-II) is 239 pc away from us
($\varpi=4.21\pm0.37$ mas), and HD~200905 (K4.5 I) is 283 pc away ($\varpi=3.59\pm 0.42$ mas).
Antares (alpha Sco, M1.5 Iab) with an estimated distance of $\approx 170$ pc does not have Gaia
parallax measurement yet. PER286 (M2.0 Ib) has an estimated distance of 4.2 kpc ($\varpi=0.20 \pm
0.04$ mas).

\section{Summary}
\label{summary}
In order to create a catalog  of stars with luminosity class I, candidate RSGs,
from Gaia DR2, we  collected 1406 bright late-type
stars with at least one spectroscopic record as class I. Spectral types were taken from the
collection by \citet{skiff16}, and in the majority of cases appeared within the uncertainty of 2
subclasses (i.e., the range of types reported for a single entry). For well known  sources, such as
those analyzed by \citet{dorda18}, \citet{negueruela16}, \citet{levesque05}, \citet{jura90},
\citet{elias85}, and \citet{humphreys78}, spectral types and luminosity classes were taken from
these works. At the present time, only a fraction equal to 13\% of this sample is known to be
associated with open clusters. For each source, we collected available photometric measurements
from 2MASS, CIO, MSX, WISE, MIPSGAL, GLIMPSE, and NOMAD catalogs and estimated their apparent
bolometric magnitudes.

We retrieved parallaxes for 1342 sources from Gaia DR2, of which 1290 have a
$(G_\mathrm{BP}-G_\mathrm{RP})$ colour. After a data filtering based on signal to noise and
astrometric quality  ($\varpi/\sigma_\varpi >4$ and $\text{RUWE} <2.7$), we were left with a
best-quality sample of 889 sources.

With the parallactic distances, we were able to estimate the stellar luminosities, and to 
build \Mbol\ versus \Teff\ diagrams of stars with different classes.

The Galactic catalog of RSGs, i.e., of very likely massive stars because of luminosity and associations with OB stars, by
\citet{humphreys78}, \citet{elias85},  \citet{jura90}, \citet{levesque05}, \citet{caron03}
contains 170 stars. 118 of these reference RSGs  had good parallaxes in DR2 and \Mbol $ < -3.6$ mag. 
While these reference RSGs appear to contain stars of class Ia, Iab (40\%) as well as class Ib (31\%),
their distribution on the \Mbol\ versus \Teff\ diagrams resembles that of class Ia, Iab,
with 81\% of them located in Area A and B. Only 44\% of class Ib stars with \Mbol $ < -3.6$ mag
fall in Area A and B.

For 609 stars (68\% of 889 analysed stars), \Mbol\ values were found smaller (brighter) than $-3.6$ mag,
with 536 of them already reported in previous literature exclusively as of classes I or II.
5\% of the them appear highly-probable massive stars (stars in Area A),
while 41 \% of them are stars in Area A and B, likely more massive than 7\Msun.

A fraction equal to $\approx 30$\% of the sample appears to be made of stars
fainter than the tip of the giant branch  (Area F).

A natural output of this luminosity exercise is a tabulated average of absolute magnitudes  of luminous late-type stars and RSGs
per spectral type. This finer grid of magnitudes will help to predict distances of extragalactic
luminous late-type stars. 

This  catalog is a little exercise on the use of accumulate spectroscopic knowledge in support of
the Gaia mission. The  catalog serve for high-resolution follow-up spectroscopy, for example, with
ongoing large spectroscopic surveys such as LAMOST and GALAH. This is important to understand the
evolution and nucleosynthesis occurring in RSGs and massive AGBs (and super-AGB stars). 
Luminosities, spectral types, and chemistry  are key ingredients for an improved study
of the Galactic structure and its recent history.

\begin{table*} 
\begin{center}
\caption{ \label{table.properties} Properties of  bright late-type stars from Table \ref{table.gaia}.}
{\tiny
\renewcommand{\arraystretch}{0.8}
\begin{tabular}{@{\extracolsep{-.06in}}rlll|rrrrr|rrrrrr|rrr}
\hline 
\hline 
Id &  {\rm Sp.Type} & Class(adopt) &  Area & { T$_{eff}$}  &{ J-K$_{\rm s}$} & { H-K$_{\rm s}$} &  A$_{\rm K_s}$(JK) & A$_{\rm K_s}$(HK) &   BC$_{K_{\rm s}}$$^a$    & K$_{\rm so}$$^b$    & M$_{\rm bol}$$^c$  &M$_{\rm bol2}$$^d$  & DM$^e$ & Mbol-Q$^f$& $V_o$ &  R \\ 
\hline 
      &                 &            &            & [K]          &  [mag]           &  [mag]             & [mag]             & [mag]          & [mag]              & [mag]          & [mag]&[mag]&[mag]& &[mag]& [$R_\odot$]\\ 
 \hline 
     1  &    M4.5     &    Ib    &     E    &3535.00 $\pm$  170.00     &    1.25     &    0.30     &   0.07 $\pm$    0.19     &   0.24 $\pm$    0.47     &    2.89     &   2.11 $\pm$    0.31     &  $-$4.35 $^{-  0.33 }  _{+  0.32 }$ &  $-$4.24 $^{-  0.26 }  _{+  0.25 }$ &   9.36 $^{-  0.10 }  _{+  0.09 }$ & 2    &   7.88    &    175   \\
     2  &      M3     &    Ib    &     F    &3605.00 $\pm$  170.00     &    1.16     &    0.28     &   0.03 $\pm$    0.01     &   0.06 $\pm$    0.03     &    2.84     &   4.27 $\pm$    0.02     &  $-$2.87 $^{-  0.10 }  _{+  0.10 }$ &  $-$2.76 $^{-  0.09 }  _{+  0.09 }$ &   9.98 $^{-  0.08 }  _{+  0.08 }$ & 2    &   9.45    &     85   \\
     3  &      M5     &    Ib    &     F    &3450.00 $\pm$  170.00     &    1.30     &    0.32     &   0.01 $\pm$    0.01     &   0.09 $\pm$    0.06     &    2.96     &   4.47 $\pm$    0.02     &  $-$2.94 $^{-  0.22 }  _{+  0.20 }$ &  $-$2.83 $^{-  0.21 }  _{+  0.20 }$ &  10.36 $^{-  0.21 }  _{+  0.19 }$ & 1    &  $..$       &     96   \\
     4  &      M2     &   Iab    &     A    &3660.00 $\pm$  170.00     &    1.06     &    0.25     &   0.22 $\pm$    0.18     &   0.24 $\pm$    0.48     &    2.80     &   1.51 $\pm$    0.29     &  $-$7.55 $^{-  0.47 }  _{+  0.43 }$ &  $-$7.54 $^{-  0.42 }  _{+  0.38 }$ &  11.86 $^{-  0.36 }  _{+  0.31 }$ & 2    &   6.39    &    716   \\
     5  &      M1     &    Ib    &     B    &3745.00 $\pm$  170.00     &    1.00     &    0.22     &  $-$0.02 $\pm$    0.22     &   0.03 $\pm$    0.70     &    2.73     &   4.29 $\pm$    0.46     &  $-$5.42 $^{-  0.59 }  _{+  0.56 }$ &  $-$5.35 $^{-  0.39 }  _{+  0.36 }$ &  12.44 $^{-  0.36 }  _{+  0.32 }$ & 2    &   9.57    &    256   \\
     6  &      K0     &   Iab    &     D    &4185.00 $\pm$   85.00     &    0.58     &    0.12     &   0.43 $\pm$    0.15     &   0.46 $\pm$    0.40     &    2.40     &   3.21 $\pm$    0.15     &  $-$4.45 $^{-  0.17 }  _{+  0.17 }$ &  $-$4.36 $^{-  0.25 }  _{+  0.25 }$ &  10.06 $^{-  0.06 }  _{+  0.06 }$ & 1    &   3.68    &    131   \\
     7  &      K3     &    Ib    &     F    &3985.83 $\pm$  170.00     &    0.72     &    0.15     &   0.24 $\pm$    0.18     &   0.23 $\pm$    0.44     &    2.55     &   3.11 $\pm$    0.29     &  $-$2.63 $^{-  0.30 }  _{+  0.30 }$ &  $-$2.58 $^{-  0.23 }  _{+  0.23 }$ &   8.30 $^{-  0.03 }  _{+  0.03 }$ & 2    &   6.02    &     62   \\
     8  &      M1     &   Iab    &     B    &3745.00 $\pm$  170.00     &    1.00     &    0.22     &   0.29 $\pm$    0.21     &   0.40 $\pm$    0.58     &    2.73     &   2.97 $\pm$    0.37     &  $-$6.03 $^{-  0.41 }  _{+  0.41 }$ &  $-$5.93 $^{-  0.29 }  _{+  0.29 }$ &  11.73 $^{-  0.16 }  _{+  0.15 }$ & 2    &   6.53    &    339   \\
     9  &      M1     &   Iab    &     B    &3745.00 $\pm$  170.00     &    1.00     &    0.22     &   0.13 $\pm$    0.18     &   0.24 $\pm$    0.43     &    2.73     &   1.75 $\pm$    0.28     &  $-$5.91 $^{-  0.36 }  _{+  0.35 }$ &  $-$5.82 $^{-  0.31 }  _{+  0.30 }$ &  10.39 $^{-  0.22 }  _{+  0.20 }$ & 2    &   5.70    &    321   \\
    10  &      M2     &     I    &     B    &3660.00 $\pm$  170.00     &    1.06     &    0.25     &   0.43 $\pm$    0.23     &   0.70 $\pm$    0.55     &    2.80     &   2.27 $\pm$    0.37     &  $-$6.18 $^{-  0.46 }  _{+  0.44 }$ &  $-$6.15 $^{-  0.37 }  _{+  0.36 }$ &  11.26 $^{-  0.26 }  _{+  0.24 }$ & 2    &   5.78    &    381   \\

\hline
\end{tabular}
}
\begin{list}{}
\item {\bf Notes.}
The identification number (Id)  from \ref{table.gaia}  is followed by the spectral type and class adopted from literature, Sp(adopt)  and Class(adopt),
by the  area occupied in the \Mbol\ vs. \Teff\ plot  (Area), the \Teff\ value, the intrinsic J-K$_{\rm s}$ and H-K$_{\rm s}$
colors, the extinction A$_{\rm K_s}$(JK) and A$_{\rm K_s}$(HK) derived from the $JK$ and $HK$ colors,  the adopted BC$_{K_{\rm s}}$,
the dereddened \Ks, K$_{\rm so}$, two estimates of bolometric magnitudes, the DM obtained with the distances of \citet{bailer18},
a flag for best near-infrared photometry (Mbol-Q),  the dereddened $V$ magnitude, $V_o$, and the stellar radius (R) estimated with the equation of \citet{josselin07}.\\ 
 A few \Aks\ values are negative. No extinction correction was applied for these stars.\\
($^a$) For \BCK, values are calculated with the formula of \citet{levesque05} and a typical error of 0.06 mag is assumed (average difference between 
the \BCK\ values of two  spectral types). \\
($^b$) The errors on the K$_{\rm so}$ values are estimated by propagating the photometric errors and the \Aks\ errors.\\
($^c$) The M$_{\rm bol}$ values are obtained with the \BCK, their errors are estimated by propagating the errors on  K$_{\rm so}$, \BCK, and DMs.\\
($^d$) The M$_{\rm bol2}$ values are obtained via integration under the SED (see Sect. \ref{calcmbol}). Errors are estimating by lowering the curve by subtracting the photometric errors,
and by lifting up the curve by adding the photometric curve. The DM error is then added by Taylor's propagation law.
($^e$) DM is here the distance module obtained with the Bailer distance. Its  error is obtained using the quoted high and low values \citet{bailer18}.\\
($^f$) Mbol-Q is set to unity  when $\varpi/\sigma_\varpi > 4$ and $\text{RUWE} < 2.7$ (889 sources), set to 2
 when $\varpi/\sigma_\varpi > 4$ and $\text{RUWE} < 2.7$ and  $JH$\Ks\  quality flags are $A$ (2MASS) or  $B$ (2MASS) or  
 $C$ (2MASS) or  $D$ (2MASS) or $M$ (HST photometry) (see Appendix). 
\end{list}

\end{center}
\end{table*}

\begin{table}
\caption{ \label{table.magbin} Magnitudes per spectral types}
\begin{tabular}{@{\extracolsep{-.07in}}rlrrrll}
\hline
Nstar& Sp.Type &  \Mbol       & \Mk     & ${\rm M_V}$  &  \Mbol-bin &  $V-K^a$\\
     &         &  [mag]       & [mag]   & [mag]        &  [mag] & \\
\hline
    3 & K0.5-K0 &   $-$5.71$\pm$   0.34    &  $-$8.12$\pm$   0.33    &  $-$6.86$\pm$   1.01    &                       $ <-$5.& 2.16     \\
    2 & K1.5-K1 &   $-$6.04$\pm$   0.22    &  $-$8.50$\pm$   0.22    &  $-$6.77$\pm$   0.27    &                       $ <-$5.& 2.29    \\
   15 & K2.5-K2 &   $-$5.74$\pm$   0.17    &  $-$8.25$\pm$   0.17    &  $-$5.97$\pm$   0.24    &                       $ <-$5.& 2.44    \\
   18 & K3.5-K3 &   $-$5.60$\pm$   0.13    &  $-$8.15$\pm$   0.13    &  $-$5.37$\pm$   0.13    &                       $ <-$5.& 2.72    \\
   20 & K4.5-K4 &   $-$5.75$\pm$   0.14    &  $-$8.35$\pm$   0.14    &  $-$5.00$\pm$   0.27    &                       $ <-$5.& 3.00    \\
   21 & K5.5-K5 &   $-$5.60$\pm$   0.11    &  $-$8.24$\pm$   0.11    &  $-$4.63$\pm$   0.18    &                       $ <-$5.& 3.70    \\
   60 & M0.5-M0 &   $-$5.72$\pm$   0.08    &  $-$8.42$\pm$   0.08    &  $-$4.31$\pm$   0.12    &                       $ <-$5.& 3.79    \\
   74 & M1.5-M1 &   $-$6.02$\pm$   0.08    &  $-$8.76$\pm$   0.08    &  $-$4.44$\pm$   0.13    &                       $ <-$5.& 3.92    \\
   91 & M2.5-M2 &   $-$6.29$\pm$   0.08    &  $-$9.09$\pm$   0.08    &  $-$4.50$\pm$   0.11    &                       $ <-$5.& 4.11    \\
   52 & M3.5-M3 &   $-$6.62$\pm$   0.13    &  $-$9.47$\pm$   0.13    &  $-$4.03$\pm$   0.15    &                       $ <-$5.& 4.58    \\
   15 & M4.5-M4 &   $-$6.40$\pm$   0.16    &  $-$9.29$\pm$   0.16    &  $-$3.58$\pm$   0.32    &                       $ <-$5.& 5.24    \\
    7 & M5.5-M5 &   $-$5.47$\pm$   0.13    &  $-$8.43$\pm$   0.13    &  $-$1.90$\pm$   0.26    &                       $ <-$5.& 6.06    \\
\hline
    3 & M6.5-M6 &   $-$4.43$\pm$   0.21    &  $-$7.52$\pm$   0.22    &     0.56$\pm$   0.93    & \text{[-3.6,-5.0]}   \\
    7 & M5.5-M5 &   $-$4.14$\pm$   0.16    &  $-$7.10$\pm$   0.16    &     0.14$\pm$   0.35    & \text{[-3.6,-5.0]}   \\
    7 & M4.5-M4 &   $-$4.12$\pm$   0.10    &  $-$7.01$\pm$   0.10    &  $-$0.95$\pm$   0.16    & \text{[-3.6,-5.0]}   \\
   13 & M3.5-M3 &   $-$4.39$\pm$   0.09    &  $-$7.23$\pm$   0.09    &  $-$0.94$\pm$   0.57    & \text{[-3.6,-5.0]}   \\
   23 & M2.5-M2 &   $-$4.55$\pm$   0.07    &  $-$7.36$\pm$   0.07    &  $-$2.10$\pm$   0.25    & \text{[-3.6,-5.0]}   \\
   19 & M1.5-M1 &   $-$4.28$\pm$   0.10    &  $-$7.01$\pm$   0.10    &  $-$2.39$\pm$   0.17    & \text{[-3.6,-5.0]}   \\
   21 & M0.5-M0 &   $-$4.43$\pm$   0.10    &  $-$7.13$\pm$   0.09    &  $-$2.99$\pm$   0.17    & \text{[-3.6,-5.0]}   \\
   17 & K5.5-K5 &   $-$4.54$\pm$   0.10    &  $-$7.18$\pm$   0.10    &  $-$2.79$\pm$   0.25    & \text{[-3.6,-5.0]}   \\
   14 & K4.5-K4 &   $-$4.57$\pm$   0.10    &  $-$7.17$\pm$   0.10    &  $-$3.93$\pm$   0.27    & \text{[-3.6,-5.0]}   \\
   31 & K3.5-K3 &   $-$4.34$\pm$   0.06    &  $-$6.90$\pm$   0.07    &  $-$3.85$\pm$   0.11    & \text{[-3.6,-5.0]}   \\
   27 & K2.5-K2 &   $-$4.29$\pm$   0.07    &  $-$6.80$\pm$   0.07    &  $-$4.09$\pm$   0.10    & \text{[-3.6,-5.0]}   \\
    5 & K1.5-K1 &   $-$4.16$\pm$   0.22    &  $-$6.63$\pm$   0.21    &  $-$4.37$\pm$   0.24    & \text{[-3.6,-5.0]}   \\
    8 & K0.5-K0 &   $-$4.19$\pm$   0.10    &  $-$6.59$\pm$   0.10    &  $-$4.35$\pm$   0.33    & \text{[-3.6,-5.0]}   \\

\hline
\end{tabular}
\begin{list}{}
\item {\bf Notes.}
Average magnitudes of stars in Table \ref{table.properties}  with  $\varpi / \sigma_\varpi > 4$ and $\text{RUWE} < 2.7$.
 The errors on the mean values are calculated as 
$\sqrt{ \frac{ \sum_{j=0}^{j=N-1} (\Mbol_j-mean)^2}{N-1}) \times \frac{1}{N}.}$
At the top, sources with \Mbol$ <-5.0$ mag and  Area A or B, or  Area C  but with secure class I from previous literature.
At the bottom, stars with $-3.6 <$ \Mbol$ <-5.0$ mag and  Area D, or E (but with secure class I from previous literature).
(a) $V-K$ colours from \citet{johnson66}. Our $V-$\Ks\ colours per spectral type
are consistent within errors with the $V-K$ colours listed in the review by
\citet{johnson66} with a mean difference of $0.26$ mag and a dispersion around the mean of $0.28$
mag. 
\end{list}
\end{table}

\begin{table}
\caption{ \label{table.magbinIaIab} Magnitudes per spectral types of stars with class Ia and Iab}
\begin{tabular}{@{\extracolsep{-.07in}}rlrrrll}
\hline
Nstar& Sp.Type &  \Mbol       & \Mk     & ${\rm M_V}$  &  \Mbol-bin &  \\
     &         &  [mag]       & [mag]   & [mag]        &  [mag] & \\
\hline
    2 & K0.5-K0 &{\bf $-$5.41$\pm$ 0.97}        &  $-$7.82$\pm$   0.96    &  $-$6.11$\pm$   0.27    &             $<-3.6$   \\
    5 & K2.5-K2 &     $-$5.6{\bf 7}$\pm$ 0.38   &  $-$8.18$\pm$   0.38    &  $-$5.86$\pm$   0.56    &             $<-3.6$   \\
    5 & K3.5-K3 &     $-$5.6{\bf 6}$\pm$ 0.52   &  $-$8.22$\pm$   0.51    &  $-$5.13$\pm$   0.56    &             $<-3.6$   \\
    3 & K4.5-K4 &     $-$5.01$\pm$   0.11       &  $-$7.61$\pm$   0.11    &  $-$3.85$\pm$   0.14    &             $<-3.6$   \\
    6 & K5.5-K5 & {\bf  $-$5.02}$\pm$   0.36    &  $-$7.65$\pm$   0.36    &  $-$3.66$\pm$   0.78    &             $<-3.6$   \\
   19 & M0.5-M0 &   $-$5.94$\pm$   0.20    &  $-$8.64$\pm$   0.20    &  $-$4.51$\pm$   0.21    &             $<-3.6$   \\
   31 & M1.5-M1 &   $-$5.80$\pm$   0.11    &  $-$8.54$\pm$   0.11    &  $-$4.09$\pm$   0.22    &             $<-3.6$   \\
   46 & M2.5-M2 &   $-$6.35$\pm$   0.14    &  $-$9.15$\pm$   0.14    &  $-$4.58$\pm$   0.23    &             $<-3.6$   \\
   21 & M3.5-M3 &   $-$7.05$\pm$   0.22    &  $-$9.90$\pm$   0.22    &  $-$4.06$\pm$   0.31    &             $<-3.6$   \\
    9 & M4.5-M4 &   $-$6.22$\pm$   0.39    &  $-$9.11$\pm$   0.38    &  $-$3.49$\pm$   0.53    &             $<-3.6$  \\
    1 & M5.5-M5 &   $-$5.33    &  $-$8.29    &  $-$2.49          &             $<-3.6$   \\
\hline
\end{tabular}
\begin{list}{}
\item {\bf Notes.}
Average magnitudes of stars in Table \ref{table.properties}  with  $\varpi / \sigma_\varpi > 4$ and $\text{RUWE} < 2.7$
and class Ia and Iab. \end{list}
\end{table}

\begin{table}
\caption{ \label{table.magbinrefRSG} Magnitudes per spectral types of reference RSGs}
\begin{tabular}{@{\extracolsep{-.07in}}rlrrrll}
\hline
Nstar& Sp.Type &  \Mbol       & \Mk     & ${\rm M_V}$  &  \Mbol-bin &  \\
     &         &  [mag]       & [mag]   & [mag]        &  [mag] & \\
\hline

  {\bf  1} & {\bf K0.5-K0} & {\bf   $-$4.68$\pm$   0.30 }   & {\bf $-$7.09$\pm$   0.28 }   &  {\bf $-$4.77 }   &             $<-3.6$   \\
  {\bf  2} & {\bf K1.5-K1} & {\bf   $-$4.43$\pm$   0.45 }   & {\bf $-$6.89$\pm$   0.44 }   &  {\bf $-$4.67$\pm$   0.33 }   &             $<-3.6$   \\
  {\bf  7} & {\bf K2.5-K2} & {\bf   $-$4.37$\pm$   0.26 }   & {\bf $-$6.88$\pm$   0.26 }   &  {\bf $-$4.51$\pm$   0.28 }   &             $<-3.6$   \\
  {\bf 10} & {\bf K3.5-K3} & {\bf   $-$5.14$\pm$   0.33 }   & {\bf $-$7.69$\pm$   0.33 }   &  {\bf $-$4.95$\pm$   0.26 }   &             $<-3.6$   \\
  {\bf  3} & {\bf K4.5-K4} & {\bf   $-$4.92$\pm$   0.46 }   & {\bf $-$7.51$\pm$   0.46 }   &  {\bf $-$4.80$\pm$   0.55 }   &             $<-3.6$   \\
  {\bf  7} & {\bf K5.5-K5} & {\bf   $-$5.33$\pm$   0.28 }   & {\bf $-$7.97$\pm$   0.28 }   &  {\bf $-$4.47$\pm$   0.35 }   &             $<-3.6$   \\
  {\bf  7} & {\bf M0.5-M0} & {\bf   $-$5.99$\pm$   0.16 }   & {\bf $-$8.69$\pm$   0.16 }   &  {\bf $-$4.80$\pm$   0.20 }   &             $<-3.6$   \\
  {\bf 15} & {\bf M1.5-M1} & {\bf   $-$6.31$\pm$   0.15 }   & {\bf $-$9.05$\pm$   0.15 }   &  {\bf $-$5.00$\pm$   0.20 }   &             $<-3.6$   \\
  {\bf 32} & {\bf M2.5-M2} & {\bf   $-$6.54$\pm$   0.12 }   & {\bf $-$9.34$\pm$   0.12 }   &  {\bf $-$4.78$\pm$   0.15 }   &             $<-3.6$   \\
  {\bf 22} & {\bf M3.5-M3} & {\bf   $-$7.19$\pm$   0.20 }   & {\bf $-$10.04$\pm$   0.20 }   & {\bf $-$4.19$\pm$   0.24 }   &             $<-3.6$   \\
  {\bf  6} & {\bf M4.5-M4} & {\bf   $-$6.57$\pm$   0.14 }   & {\bf $-$9.46$\pm$   0.14 }   &  {\bf $-$3.43$\pm$   0.25 }   &             $<-3.6$   \\
  {\bf  1} & {\bf M5.5-M5} & {\bf   $-$5.55$\pm$   0.49 }   & {\bf $-$8.51$\pm$   0.35 }   &  {\bf $-$2.19 }   &    $<-3.6$    \\
    
\hline
\end{tabular}
\begin{list}{}
\item {\bf Notes.}
Average magnitudes of stars plotted in the upper panels of Fig. \ref{compare_ref}. 
\end{list}
\end{table}

\appendix{
\section{A. Notes on photometric data}

Typically, initial coordinates by \citet{skiff16} are good to within a few arcseconds. A few
coordinates were corrected with SIMBAD. An iterative process was needed to make sure to properly
identify the counterparts at different wavelengths. The Galactic plane is crowded with sources. 

For stars at longitude $|l| >1$\degr\ and latitude $|l| >0.5$\degr, measurements were associated
automatically with a selection of good flags to ensure quality. MSX upper limits measurements were
discarded, and WISE sources were chosen with a minimum signal-to-noise larger than 2. GLIMPSE
matches were associated with a magnitude cut at 10 mag, and when a WISE source was existing
positional coincidence was inspected. The searched stars were usually the brightest at near- and
mid-infrared wavelengths, and chart identification was easy. 2MASS matches are as in the WISE and
GLIMPSE catalogs. Due to saturation and centroid problems, a few 2MASS identifications had to be
fixed (e.g. BD+54~315, VY~CMa, Cl*~Westerlund~1~26, MZM29, MZM33, RSGC1-F08, IRAS~17433$-$1750).
For stars HD~126152, HD~149812, HD~227793, BD~+36~4025, which have good quality parallaxes but no
2MASS errors, we assumed an error in \Ks=0.8 mag (see the quality flag provided in Table
\ref{table.properties}). For omi02~Cyg, $JK$ photometry was taken from \citet{morel78}. For stars
$[$MMF2014$]$~78, $[$MFD2010$]$~5, $[$GLIMPSE9$]$-6, and $[$MMF2014$]$~46/$[$MFD2010$]$~8, HST $HK$
data were available \citep{messineo10}; for the faint OGLE~BW3~V~93508, near-infrared magnitudes are
from \citet{lucas08}. For the highly-crowded central region ($|l|<1.0$\degr and $|b|<0.5$\degr),
only the $K$-band photometry of \citet{liermann09} is provided, and for stars IRC$-$30320 ,
IRC$-$30322, $[$RHI84$]$~10$-$565, MZM115 the 2MASS photometry. For LHO036, which as a parallax,
additional $JH$ measurements taken from the work of \citet{stolte15}.

Matches were confirmed with a visual inspection of 2MASS and WISE images, as well as of their SEDs.
After the visual inspection, a few measurements were discarded as of poor quality (e.g., confused,
highly saturated, or strong background emission) and not compatible with the SED. For stars
$[$MMF2014$]$~46, GLIMPSE9-6, RSGC2-8, RSGC2-14, 2MASS~J18451760-0343051, and
2MASS~J18451722-0343136, MSX matches were removed. For stars Cl*~Westerlund~1~20,
Cl*~Westerlund~1~75, $[$MMF2014$]$~46, GLIMPSE9-6, RSGC1-F08, RSGC1-F05, and RSGC1-F01, WISE matches
were removed because they are blended with other sources. For stars $[$HSD93b$]$~48,
$[$MNG2014$]$~vdB-H~222~778, $[$MNG2014$]$~vdB-H~222~664, $[$MNG2014$]$~vdB-H~222~479,
$[$MMF2014$]$~78, 2MASS~J18410261$-$0552582, HD~195214, and 2MASS~J18392955$-$0544222 only $W4$ 
measurements were removed because sources were too faint or confused at this longer wavelength. For
stars 2MASS~J17361839-2217306, RSGC1-F07, RSGC1-F10, RSGC1-F03, 2MASS~J18395282-0535172, both
$W3$ and $W4$ magnitudes were discarded. For HD~14580 and Cl*~Westerlund~1~26, $W1$ and $W2$
magnitudes did not fit their SED.

\begin{acknowledgements}
This work has made use of data from the European Space Agency (ESA) mission {\it Gaia}
(\url{http://www.cosmos.esa.int/gaia}), processed by the {\it Gaia} Data Processing and Analysis
Consortium (DPAC, \url{http://www.cosmos.esa.int/web/gaia/dpac/consortium}). Funding for the DPAC
has been provided by national institutions, in particular the institutions participating in the {\it
Gaia} Multilateral Agreement. This publication makes use of data products from the Two Micron All
Sky Survey, which is a joint project of the University of Massachusetts and the Infrared Processing
and Analysis Center/California Institute of Technology, funded by the National Aeronautics and Space
Administration and the National Science Foundation. This work is based on observations made with
the Spitzer Space Telescope, which is operated by the Jet Propulsion Laboratory, California
Institute of Technology under a contract with NASA.
This research made use of data products from the Midcourse Space Experiment, the processing of which
was funded by the Ballistic Missile Defence Organization with additional support from the NASA
office of Space Science. This publication makes use of data products from WISE, which is a joint
project of the University of California, Los Angeles, and the Jet Propulsion Laboratory/California
Institute of Technology, funded by the National Aeronautics and Space Administration. 
This work makes use of the Naval Observatory Merged Astrometric Dataset (NOMAD). This research
has made use of the VizieR catalogue access tool, CDS, Strasbourg, France, and SIMBAD database.
This research has made use of NASA’s Astrophysics Data System Bibliographic Services. 
We thank the anonymous referee for his/her very constructive comments.
This work was partially supported by the National Natural Science Foundation of China
(NSFC-11773025, 11421303), and USTC grant KY2030000054. 
\end{acknowledgements}

\end{document}